\documentclass[utf8]{frontiersSCNS}
\usepackage{url,hyperref,lineno,microtype,subcaption}
\usepackage[onehalfspacing]{setspace}
\newcommand{\RomanNumeralCaps}[1]
    {\MakeUppercase{\romannumeral #1}}
\usepackage{xcolor}

\def\keyFont{\fontsize{8}{11}\helveticabold }
\def\firstAuthorLast{Ziemer {et~al.}} 
\def\Authors{Tim Ziemer\,$^{1,*}$, Fabian Wetjen\,$^{1}$ and Alexander Herbst\,$^{1}$}
\def\Address{$^{1}$Bremen Spatial Cognition Center, University of Bremen, Bremen, Germany}
\def\corrAuthor{Tim Ziemer}

\def\corrEmail{ziemer@uni-bremen.de}

\begin{document}
\onecolumn
\firstpage{1}

\title[Mosquito Antenna Base]{The Antenna Base Plays a Crucial Role in Mosquito Courtship Behavior} 

\author[\firstAuthorLast ]{\Authors} 
\address{} 
\correspondance{} 

\extraAuth{}

\maketitle

\begin{abstract}

\section{}
Mosquitoes are vectors of diseases like malaria, dengue fever, yellow fever, chikungunya and Zika. For mosquito control it is crucial to understand their hearing system, as mosquitoes' courting behavior is mostly auditory. Many nonlinear characteristics of the mosquito hearing organ have been observed through behavioral studies and neural measurements. These enable mosquitoes to detect and synchronize to other mosquitoes. Many hypotheses concerning the role of the flagellum and the fibrillae of the antenna in mosquito hearing have been made, and neural processes have been considered as the origin of the nonlinearities. In this study we introduce a geometric model based on the morphology of the mosquito antenna base. The model produces many of the observed nonlinear characteristics, providing evidence that the base of the antenna plays a crucial role in mosquito hearing. Even without neural processing, the antenna response to sound produces behaviorally relevant cues that can inform about the presence, location and sex of other mosquitoes.

\tiny
 \keyFont{ \section{Keywords:} mosquito hearing, acoustics, courtship behavior, antenna, malaria, dengue fever, yellow fever, vector control} 
\end{abstract}

\section{Introduction}
Mosquitoes are amongst the deadliest animals on earth, as they are vectors of diseases like malaria, dengue fever, yellow fever, chikungunya and Zika, that cause over $700,000$ deaths per year and inflict suffering on more than $50$ million people \citep{who,surveillance}. However, out of over $3500$ extant species only around $100$ are invasive \citep{species}. Still, $40$ to $80$\% of the global population are at risk of one or more major vector-borne diseases (VBD) \citep{surveillance,rearing}.

When regional proliferation of a dangerous mosquito species has been identified, interventions by means of insecticide spraying, Sterile Insect Technique (SIT) or Incompatible Insect Technique (IIT) can be carried out \citep[Chap. 20]{buch} to protect the population: In the SIT \citep{sit}, male mosquitoes are sterilized, e.g., by means of x-ray radiation, and then released into the wild. Mating with sterile mosquitoes decrease the females’ reproductive potential, which can ultimately reduce the population. In the IIT male mosquitoes are infected with Wolbachia bacteria that partly sterilize male mosquitoes, leading to an effect similar to SIT. A third option is the release of insects with a dominant lethal (RIDL) gene. This lethal gene is female-specific, so all progenies are male.

To date, mosquitoes are manually identified and counted in a large network of measurement stations \citep{surveillance}, which is extremely inefficient and costly. Acoustical sensors \citep{dinarte,goodit,lyn} offer an efficient alternative. They automatically classify and count mosquito species and sex. Unfortunately, the transfer from the lab to real world conditions is not robust so far.

As sound is assumed to be the only sexual stimulant in some mosquito species \citep{erect}, and mosquitoes identify, recognize, localize and approach other mosquitoes ``seemingly unerringly'' \citep{diseasesound} by means of sound \citep{tischner,Jackson16734,current,sciencesongs,f0sync,tunedis,active,simuldoubl}, understanding their auditory system may help us automatically classifying and counting mosquito species and sex in the future and, eventually, understand and control mosquitoes' courtship behavior.

In this paper we introduce and evaluate a biologically-inspired model of the mosquito antenna base. The model is an analysis-by-synthesis approach that enables us to find out more about the nonlinear transduction mechanisms inside the mosquito antenna that enables them to detect, identify and localize other mosquitoes. 

\section{Mosquito Hearing}
\begin{figure}
	\centering
	\includegraphics[width=2.2in]{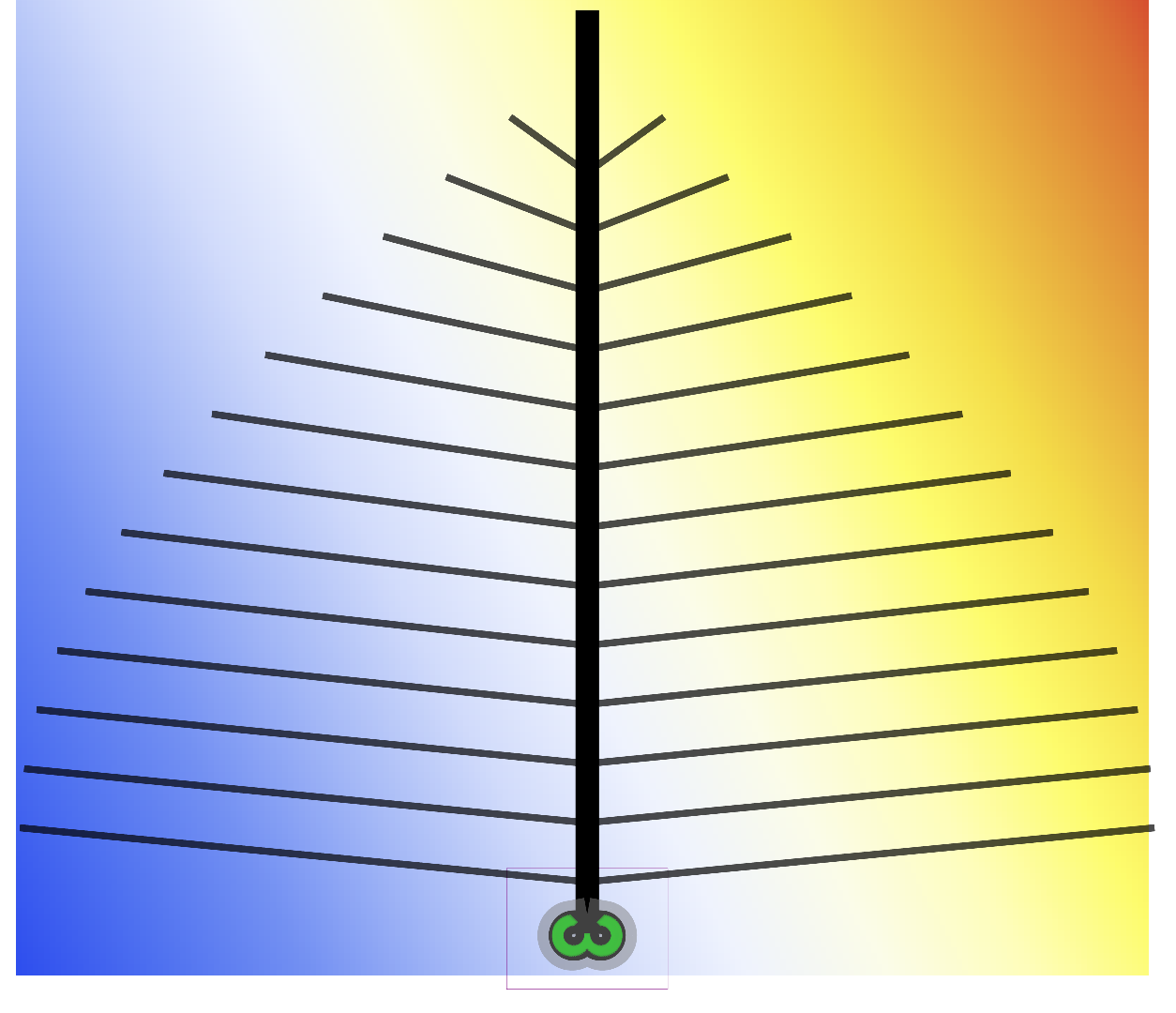}
	\includegraphics[width=2.2in]{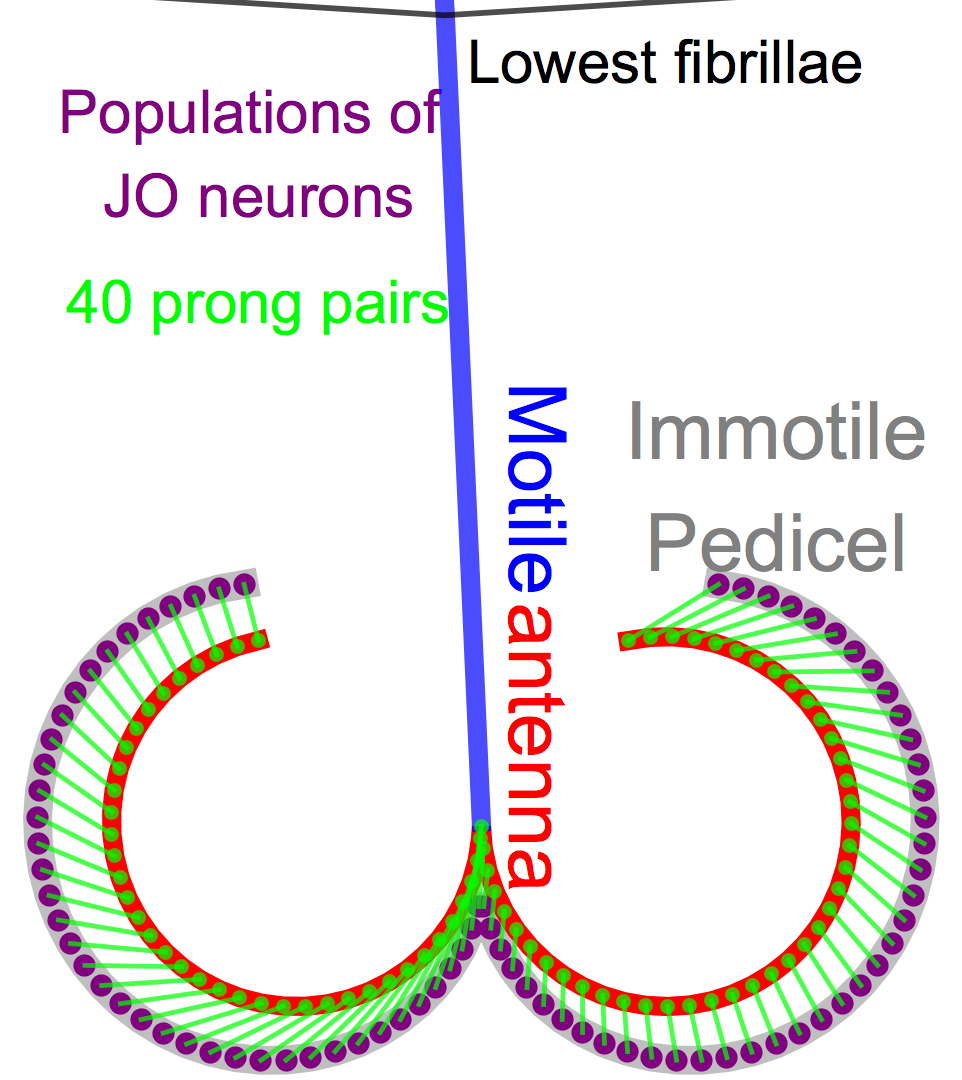}
	\includegraphics[width=2.2in]{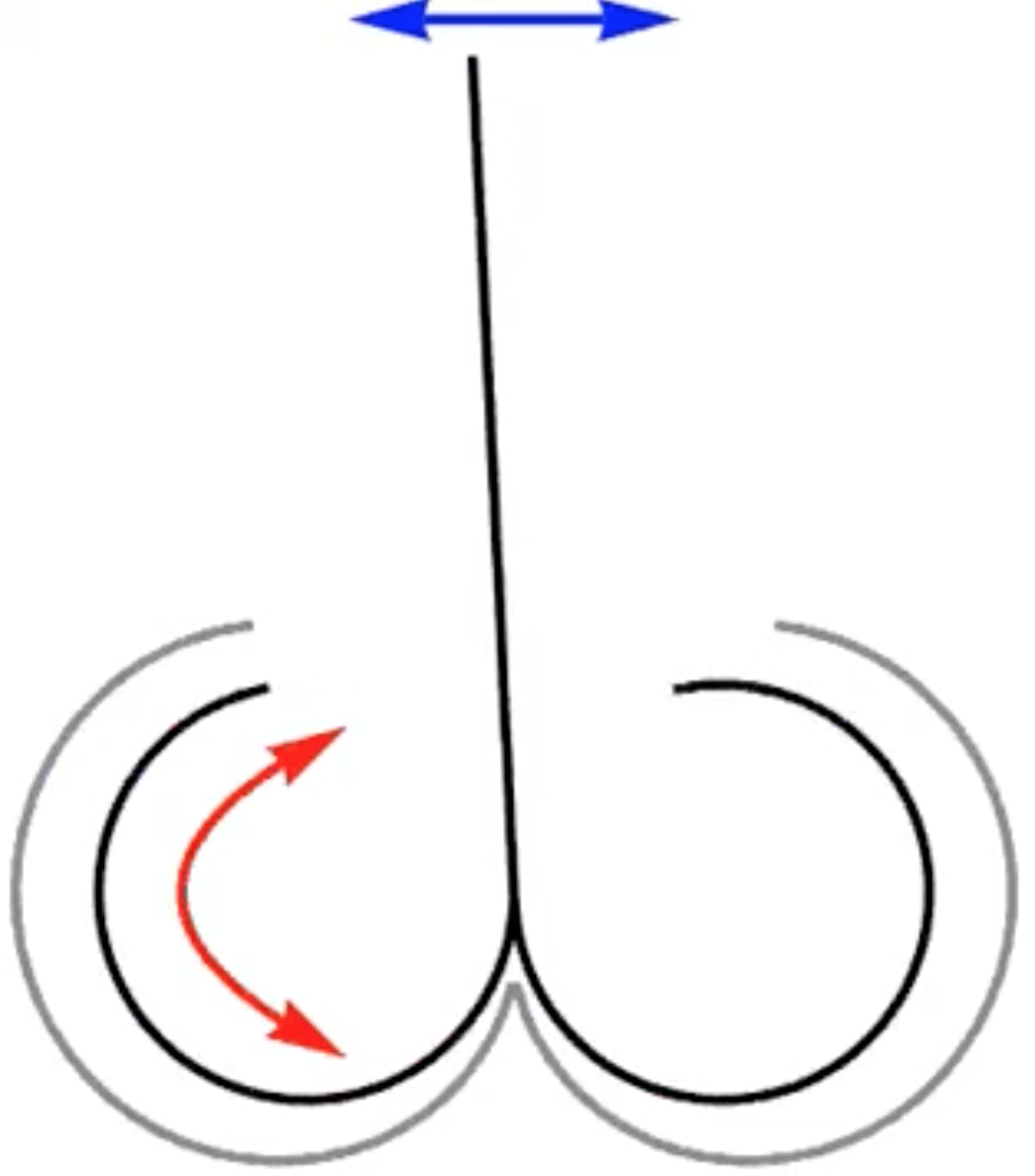}
	\caption{\label{fig:base}Mosquito antenna in a sound field (left). The antenna base (center) contains the linear (blue) and curved (red) part of the motile antenna. The curved part contains prongs (green dots) which are connected (green lines) to populations of neurons (purple dots) on the Johnstons's organ (JO), which is embedded in the immotile pedicel (gray). The model in motion can be seen on \url{https://youtu.be/D4yNbGvX4lU}. Most importantly, the model produces a transfer from deflection to rotation (blue and red arrows on the right).}
\end{figure}

Mosquitoes are amongst the best-hearing insects \citep{besthearing}. The periphery of the mosquito hearing organ is the antenna pair. The morphology of the antenna is described, e.g., in \citep{spontaneous,tischner,structure,schwartzkopff} and depicted in \citep{mechan,spontaneous,stiffness}. As can be seen in Fig. \ref{fig:base} (left), an antenna consists of a flagellum (bold black line) to which fibrillae of multiple lengths are attached. The base of the flagellum is encapsulated by the immotile, bulbous pedicel, which contains the Johnston's Organ (JO).

In the JO mechanoreceptors sense particle velocities \citep{Davis1976}. This helps for hearing sound sources. But the JO is also a somatosensory system, as the mosquito may detect wind as well as its own flight tone, which are necessary for navigation. Furthermore, the JO contains thermoreceptors and different types of chemoreceptors. Consequently, the $15,000$ neurons in the JO are not exclusively auditory \citep{farfield}.

Auditory neurons in the JO are tuned to frequencies between $85$ to $470$ Hz \citep{Lapshin3927} and mechano-electrical transduction is limited to frequencies below $400$ Hz. This is true for both female and male mosquitoes \citep{tunedis}. When flying, the mosquito's wing beats produce a flight tone, i.e., a harmonic spectrum with a fundamental frequency that roughly lies between $200$ Hz and $1200$ Hz, \citep{chen,lyn}. The fundamental frequency mainly depends on species and sex but can be influenced by flight maneuvers, age, humidity, temperature or diseases. 
Throughout most species, the male mosquito is considerably smaller, creating a much higher fundamental frequency. Note that mosquitoes are very small compared to the wavelengths that their wing beats produce, so they can be considered as point sources whose sound energy largely remains within their near field rather than propagating to the far field. Consequently, mosquitoes seem to hear each other only over a range of some centimeters \citep{comparative,f0sync,sciencesongs,farfield}.

The mosquitoes' frequency tuning implies that many mosquitoes cannot even directly hear male mosquitoes, as males exclusively create inaudibly high frequencies \citep{tischner}. However, during mate attraction, mosquitoes of the same species but opposite sex \emph{synchronize} \citep{sciencesongs}. Synchronization is also referred to as \emph{tune-in} \citep{tunedis}, \emph{harmonic convergence} \citep{sciencesongs} and \emph{frequency locking} \citep{2012-HarmonicSynchronization} in the mosquito literature, and as \emph{lock-in} or \emph{entrainment} in other domains (e.g., \citep{ice,noorden}). Every second wing beat of the female mosquito is synchronous to every third wing beat of the male mosquito, so what matches are the second partial of the male mosquito and the third partial of the female mosquito. In musical terms, they produce a fifth, not a unison. Mate attraction is bi-directional, i.e., both male and female mosquitoes modulate their flight tones to create the fifth \citep{sciencesongs}, which tends to lie above $1$ kHz. It is assumed that sound is the \textcolor{red}{major} sexual stimulant in many mosquito species \citep{erect}, as courtship and copulation only takes place between flying mosquitoes and not, if either of the mosquitoes is at rest \citep{erect,distortion}\cite[p. 324]{insect}\cite[p. 20]{buch}
. At the same time, even in absence of female mosquitoes, male mosquitoes have been found to copulate with loudspeakers if they play the right sound \cite[p. 57]{insect}.Note that the synchronization alters between perfect and imperfect match, creating beats every now and then.

As mosquitoes tend to synchronize at a frequency that they cannot even hear directly, \cite{tunedis} hypothesize a ``nonlinear interaction between her own flight tones and those of a nearby male''. This hypothesis is supported by the finding that most auditory neurons in the JO are tuned to $190$ to $270$ Hz \citep{Lapshin3927} and mosquitoes are most sensitive to frequencies below $200$ Hz \citep{tunedis}. This frequency region lies below most emitted mosquito sounds but rather covers the \emph{difference frequency} between the harmonics of two wing beat sounds. We also refer to the difference frequency as the \emph{beat frequency}. This frequency difference can refer to the fundamental frequency of a male and a female mosquito of the same species. Mosquitoes could hear the fundamental frequency difference of their own and a nearby wing beat. However, it has been found that mosquitoes also synchronize to the recording of an opposite sex mosquito when the fundamental frequency has been removed, and even when only a pure tone near their synchronization frequency is being played \citep{sciencesongs}. So another reason for the frequency tuning seems to be that it corresponds to the frequency difference of their overtones in the initial state of their synchronization progress during mating, i.e., to the beat frequency. During synchronizing, the beat frequency may fluctuate between zero and some hundred Hertz. In both cases nonlinearities in the auditory system are necessary to produce a difference frequency, i.e., to represent the beat frequency.

The fact that mosquitoes approach not only other mosquitoes, but even sound sources that produce an inaudible high pure tone near their preferred synchronization frequency \cite{sciencesongs} is evidence that the difference frequencies enable mosquitoes to localize sound sources.

Another nonlinearity that has been observed is that periodic oscillations in the JO as well as acoustically evoked field potentials often exhibit a frequency doubling compared with the stimulation frequency \citep{distortion,simuldoubl,Lapshin3927,tischner,invivowindmill}. The doubled frequency can even be stronger than the original stimulation frequency. The intensity and the number of harmonic frequencies depends on the intensity of the input signal \citep{distortion}. However, the strength of the doubling is different between different prongs \citep{wishart}. So in addition to difference frequencies, harmonic distortion seems to take place, in which the doubled frequency may dominate the biomechanical and neural response at least at some locations along the JO.

It has been observed that some axons of the JO sensory unit respond in anti-phase to each other  \citep{Lapshin3927}. This is likely to happen because mechanical motion of the antenna creates anti-phase excitation at the JO, so the axons respond in anti-phase in order to amplify the excitation.


The origin of such nonlinearities that enable mosquitoes to synchronize is still an open question, but understanding the mechanisms underlying the selection of conspecific mates ``(\ldots) could open the door to novel ways of attracting or trapping and killing males or females'' \cite{mating}.  
Our model sheds light on the nature of the nonlinear interaction in mosquito hearing, beginning in the mosquito antenna base, as ``(f)uture research is needed to resolve these questions, but in either case, it seems, the sensory periphery will take a centre-stage role.''\cite{comparative}


In summary, a number of nonlinearities has been observed in the mosquito antenna:

\begin{enumerate}
    \item harmonic distortions
    \begin{enumerate}
    \item that can even be stronger than the input frequency
    \item whose number and intensity depend on the prong location
    \item whose number and intensity depend on the intensity of the input signal
    \end{enumerate}
    \item anti-phase responses
    \item difference frequency between female and male $f_0$
    \item difference frequency near the synchronization frequency
    \item difference frequencies provide localization information
\end{enumerate}

We use these $5$ observations as criteria to evaluate to what degree the antenna base contributes to the nonlinearities that enable mosquitoes to detect, localize and synchronize to other mosquitoes.

\section{Related Work}
\cite{mechan} model the antenna mechanics as a forced-damped oscillator. However, the model simplifies the curved septa as a straight line instead of sticking to the actual geometry of the mosquito antenna morphology, and it has physically implausible properties \citep{thesis}, so the model needs some further explorations to be explanatory for the nonlinear processing in the mosquito antenna.

\cite{stiffness} model an antenna flagellum using a finite element method and conclude that the main function of the varying stiffness along the length of the flagellum could serve for mechanical frequency selectivity, i.e., low pass and bandpass filtering. This is in agreement with the observations of \cite{modelseries} who model the flagellum as a damped harmonic oscillator with a series of filters.

\cite{review} reviews a number of additional insect ear models that mostly aim at exploring active amplification through biophysical and neurobiological considerations.

All of these models concentrate on physical properties of the flagellum and partly assume additional nonlinearities in the neural mechanisms of mosquito hearing. The model in \cite{distortion} goes one step back and considers mostly the rotation of the basilar plate relative to the pedicel. With only little additional filtering, their model produces the difference frequency of a two-tone input, as well as harmonic distortions. These are two of the above-listed criteria. Our model does the same but sticks much closer to the morphological geometry of the mosquito antenna to be able to meet all of the above-listed criteria.

\section{Mosquito Antenna Model}
Our antenna model is a geometric model based on \citep{jasaabstract}. As a simplification we neglect physical properties, like mass, Young's modulus, sound impedance, restoring forces, etc. Instead, we consider all parts of the antenna as perfectly rigid and the connection between the motile and immotile part of the antenna as perfectly flexible. This is a reasonable simplification, as the antenna is a structure that is very small compared to the wavelengths that is receives and stiff compared to the surrounding air, so the ``flagellum moves like a stiff rod rocking about its socket when stimulated at frequencies below or around the best frequency'' \cite{besthearing}. The model approximates the morphological relations of the antenna as described, e.g., in \citep{structure,tischner}. Our model focuses on the antenna base.%

The model is illustrated in Fig. \ref{fig:base} (center). It assumes a motile flagellum (blue) whose deflection equals the particle deflection of the incoming sound waves. At the distal end of the scape, the antenna flagellum is extended by the curved septa (s, red)
\begin{equation}
s (\phi) = \left| \cos\left(\phi+d(t)\right)\right|, ~\frac{3\pi}{4}\leq\phi\leq\frac{9\pi}{4} \ .
\end{equation}
Here, $d(t)$ is the deflection of the antenna, which is proportional to the particle deflection caused by sound waves. Together, the flagellum and the septa comprise the motile antenna part. The flagellum and the septa meet at the socket, the basilar plate at $\phi=\pi$. The motile septa contains $40$ prong pairs (green dots) at %
\begin{equation}
\phi_{s,i} (t) = d(t)+\frac{3 \pi}{4}+i \frac{6 \pi}{4\times79}, ~i=0, \ldots, 79 \ .
\end{equation}
These are connected to auditory nerve cells of the Johnston's organ (JO, purple) via the attachment cells of the scolopidia (green lines), which are located at the discrete positions
\begin{equation}
\phi_{p,i} = \frac{3 \pi}{4}+i \frac{6 \pi}{4\times79}, ~i=0, \ldots, 79 
\end{equation}
along the immotile pedicel
\begin{equation}
p (\phi) = 0.2 + \left| \cos\left(\phi\right)\right|, ~\frac{3\pi}{4}\leq\phi\leq\frac{9\pi}{4} \ .
\end{equation}
We consider the pedicel immotile and perfectly rigid, so the rotation of the septa is restricted to $\frac{\pi}{4} <d(t)<\frac{3\pi}{4}$. From a macroscopic viewpoint, our antenna model describes a transform from deflection (blue arrow) to rotation (red arrow) as depicted in Fig. \ref{fig:base} (right). However, as illustrated in Fig. \ref{fig:base} (center), the model imposes a one-to-many transform: the deflection $d(t)$ of the flagellum creates $80$ rotations, one at each prong. These rotations stretch and compress the attachment cells of the scolopidia at $80$ locations like
\begin{equation}%
r_{i}(t) = \left|\left|p (\phi_{p,i}) -s (\phi_{s,i}(t))\right|\right|_2
\end{equation}
which is the Euclidean distance between the septa and the pedicel at the locations of the $80$ prongs. These are functions of time because the septa moves as a function of time.

So the input of the model is a continuous function $d(t)$ or a discrete time series $d[\tau]$ that represents the flagellum deflection caused by an incoming sound wave. Likewise, the output can be the continuous functions $r_i(t)$ or discrete time series $r_i[\tau]$. We use capital letters to indicate that we are dealing with the frequency spectrum after a Discrete Fourier Transform ($\text{DFT}$), i.e.,

\begin{equation}
D(\omega) = \text{DFT}\left[d(t)\right]
\label{eq:dft1}
\end{equation}
and
\begin{equation}
R_i(\omega) = \text{DFT}\left[r_i(t)\right] \ .
\label{eq:dft2}
\end{equation}

In nature, the antenna is deflected by sound waves. The wing beats of the mosquito itself creates a flight tone that propagates to its antenna pair. Furthermore, the flight tones of other mosquitoes travel to the antenna. Consequently, the antenna is deflected by superposition of the mosquito's own wing beat sound, the wing beat sound of other mosquitoes and additional ambient sounds.

In the mosquito the stretch and compression of the attachment cells open channels that cause neural activity in the auditory part of the antennal nerve \citep{Lapshin3927}. However, the neural encoding and further processing is out of scope of this paper.

\section{Evaluation}
The evaluation of our mosquito antenna model is explorative and qualitative. We examine which of the $5$ nonlinear characteristics of mosquito hearing is reproduced by the model.

As the deflection of the antenna $D(\omega)$ and the stretch and compression of attachment cells take place at different orders of magnitude, we normalize amplitudes as

\begin{equation}
D(\omega) [\text{dB}] =   20 \left( \log_{10}\frac{D(\omega)}{\text{max}\left|D(\omega)\right|}\right)
\end{equation}
and 
\begin{equation}
R_i(\omega) [\text{dB}] =   20 \left( \log_{10}\frac{R_i(\omega)}{\text{max}\left|R_i(\omega)\right|}\right) \ ,
\end{equation}
respectively. Due to this normalization, all amplitudes take values $\leq 0$ dB, where $D(\omega)=0$ dB refers to the loudest input frequency and $R_i(\omega)=0$ dB to the strongest response frequency at the $i$th prong.

\subsection{Harmonic Distortion}
\begin{figure}
	\centering
	\includegraphics[width=2.1in]{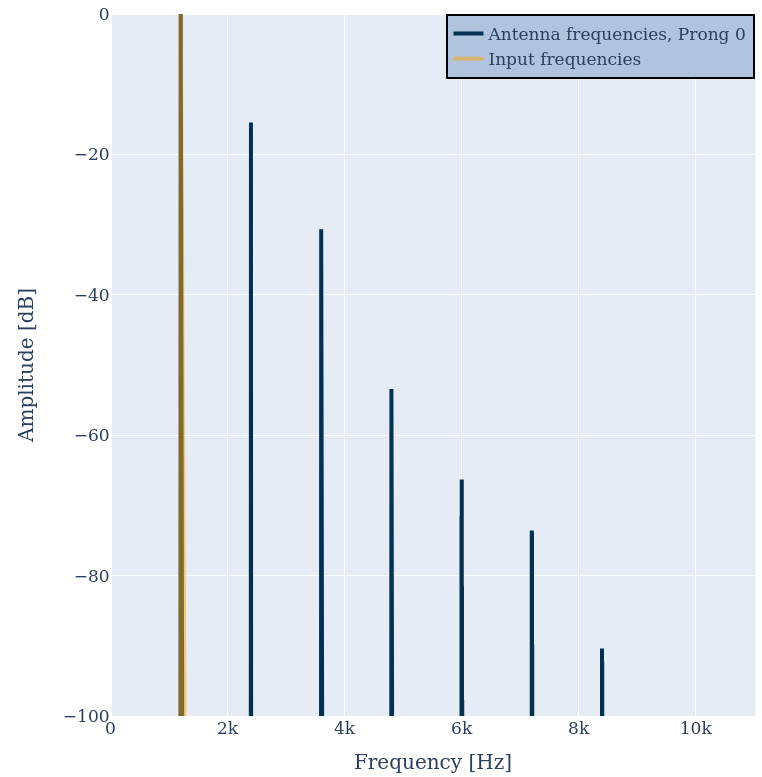}
	\includegraphics[width=2.1in]{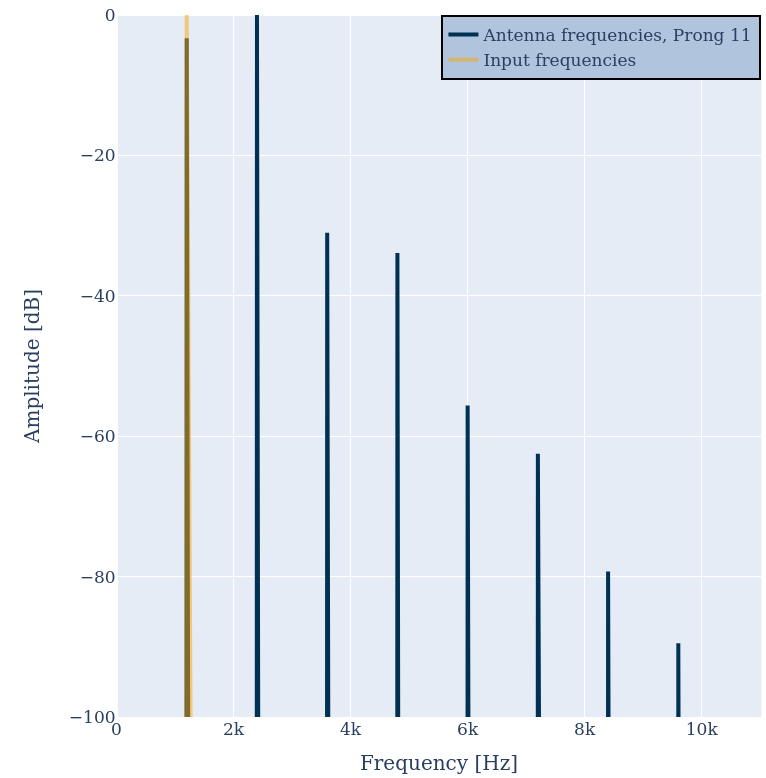}
	\includegraphics[width=2.1in]{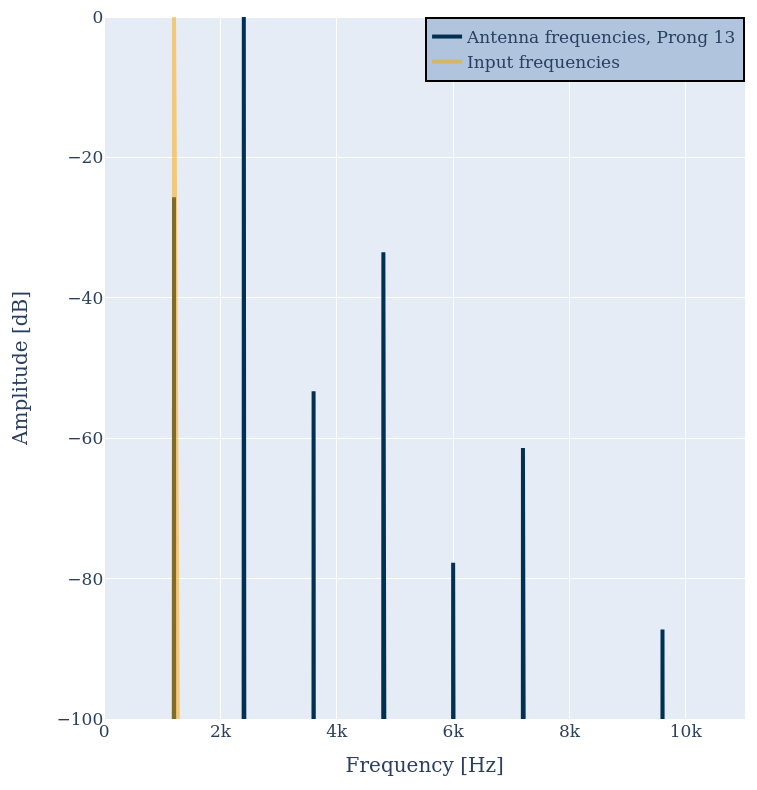}
	\caption{\label{fig:1200_hz} Antenna response $R_0(\omega)$ (blue, left) and $R_{11}(\omega)$ (blue, center) to a single frequency of $1200$ Hz (orange). The output spectrum either peaks at the input frequency or the double frequency. Additional harmonic overtones are visible. When reducing the amplitude of the input by a factor of $2$ (right) the distribution of distortion frequencies in $R_{11}(\omega)$ changes and the distortion factor reduces.}
\end{figure}

First we input a pure tone with a frequency of $1200$ Hz to the model. Figure \ref{fig:1200_hz} illustrates the frequency spectrum of the input signal $D(\omega)$ (orange) and the antenna response $R_i(\omega)$ (blue) of prongs number $0$ (left) and $11$ (center). In both cases the antenna responds with the input frequency and additional harmonic overtones, as suggested by criteria 1. Harmonic distortion can be observed at all prongs and always contains even and odd integer multiples of the incoming frequency. However, the amplitude distribution is different for different prongs, as suggested by criteria 1b. At prong $11$ the first overtone has an even larger amplitude than the fundamental frequency. This is in agreement with the frequently observed frequency doubling \citep{distortion,Lapshin3927,tischner} evident at different prongs, and with criteria 1a. When reducing the amplitude of the input spectrum $D(\omega)$ by a factor of $2$ (center vs. right) the distribution of harmonic distortion frequencies in $R_{11}(\omega)$ changes and the overall distortion factor reduces. This is in agreement with criteria 1c.

\subsection{Anti-Phase Response}
\begin{figure}
	\centering
	\includegraphics[width=5in]{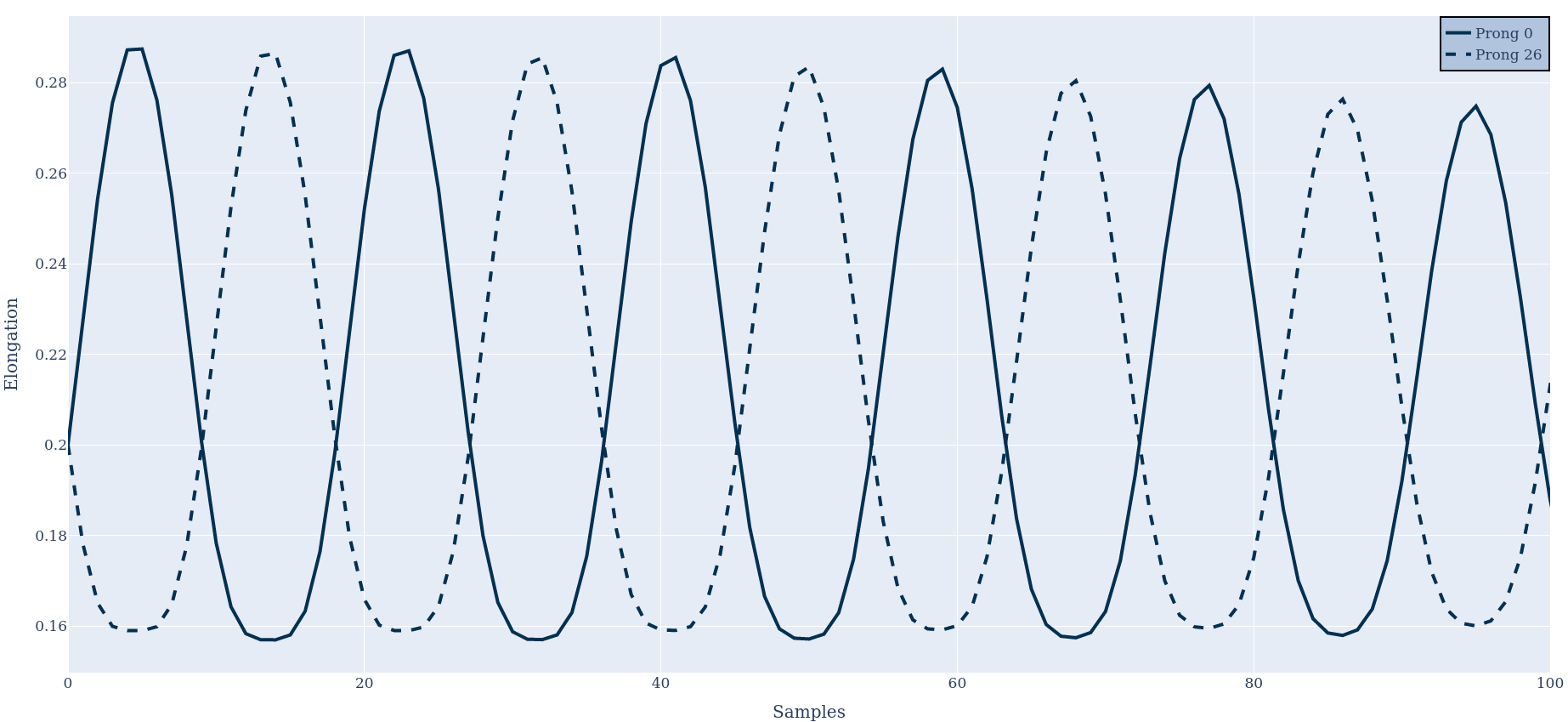}
	\caption{\label{fig:antiphase} The antenna response in the time domain at prongs 0 and 26, when stimulated with two frequencies of $1200$ Hz and $1240$ Hz.}
\end{figure}

Figure \ref{fig:antiphase} shows the antenna response in time domain at two different prongs.  It can be observed that the two prongs respond in anti-phase to each other, as suggested by criteria 2.

\subsection{Difference Tones}
\begin{figure}
	\centering
	\includegraphics[width=2.2in]{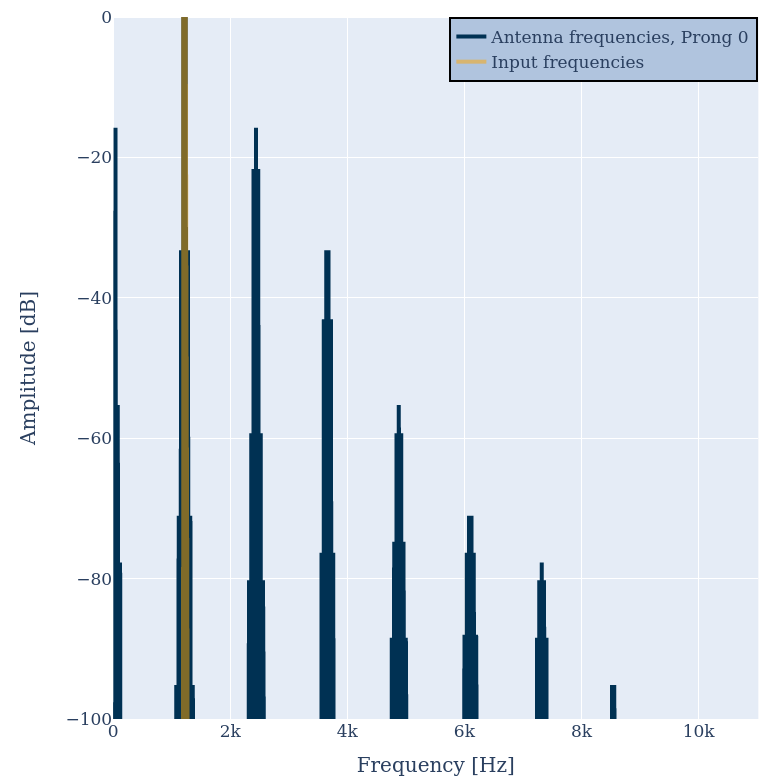}
	\includegraphics[width=2.2in]{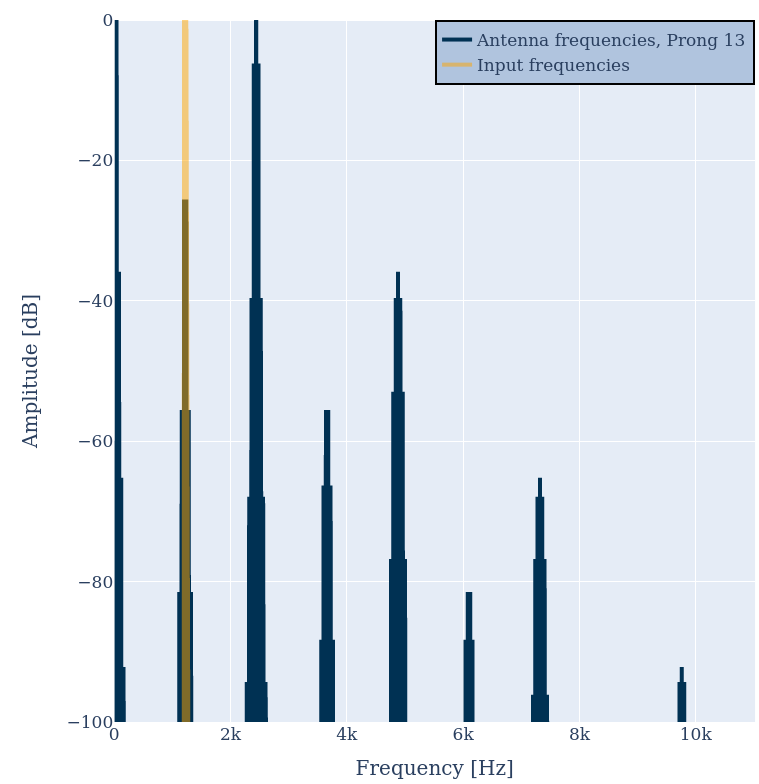}
	\includegraphics[width=2.2in]{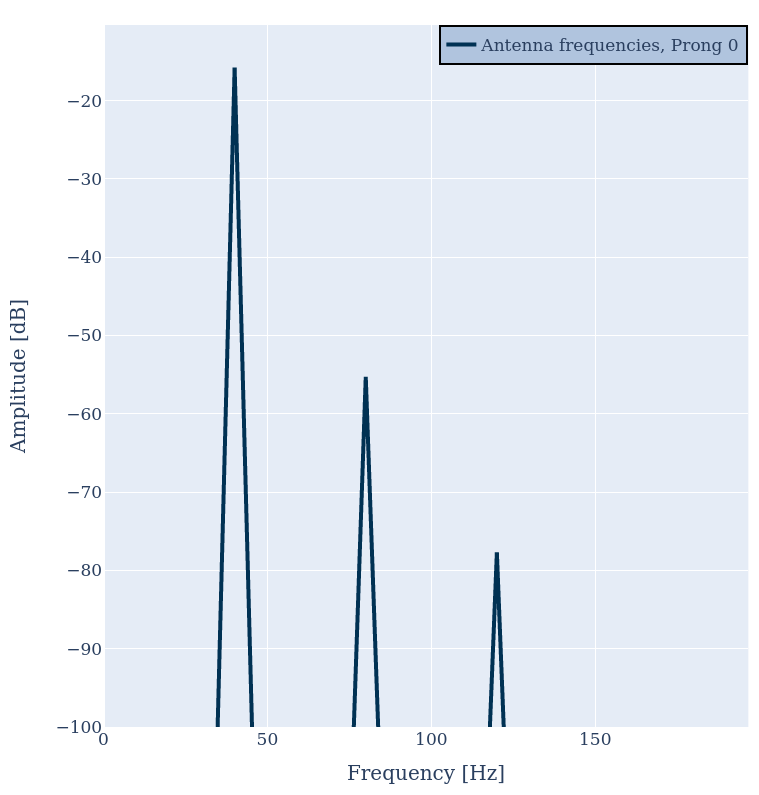}
	\caption{\label{fig:input_amplitude} Antenna outputs $R_0(\omega)$ (left) and $R_{13}(\omega)$ (center), when evoked by two proximate frequencies of $1200$ and $1240$ Hz. The highest amplitude can be found at the input frequencies, their octaves, or their difference frequency. Zooming into the low-frequency region (right) shows the beat frequency plus harmonic distortion.}
\end{figure}

Note that exciting the antenna with a single frequency simulates a very simple scenario in which a passive mosquito is exposed to a sine wave. To simulate flight, we input a superposition of two proximate pure tones of $f_\text{int}=1200$ and $f_\text{ext}=1240$ Hz to the model. One of these frequencies simulates a frequency from the wing beat of the modeled mosquito (internal) the other, of another mosquito nearby (external). The exemplary result of two different prongs is illustrated in Fig. \ref{fig:input_amplitude}. The same behavior concerning the strong representation of the input frequencies in the output time series can be observed. Again, the number and amplitude of distortion products depends on the prong number (cf. $R_0(\omega)$, left and $R_{13}(\omega)$, center). The amplitude of the double frequencies can even be stronger than the amplitude of the two input frequency.

In addition, a low-frequency component is visible at most prongs. In the prong output, these low-frequency components can even be stronger than the amplitudes of the original input frequencies, as illustrated in Fig. \ref{fig:input_amplitude}  for $R_{13}(\omega)$ (center). As can be seen in  Fig. \ref{fig:input_amplitude} (right), the low frequency components are the difference frequency of the two input frequencies, i.e., $f_\text{beat}=f_\text{ext}-f_\text{int}=40$ Hz, which is in accordance with criteria 4. In addition, harmonic distortion can be observed, i.e., $80$ Hz and $120$ Hz.

\begin{figure}
	\centering
	\includegraphics[width=3in]{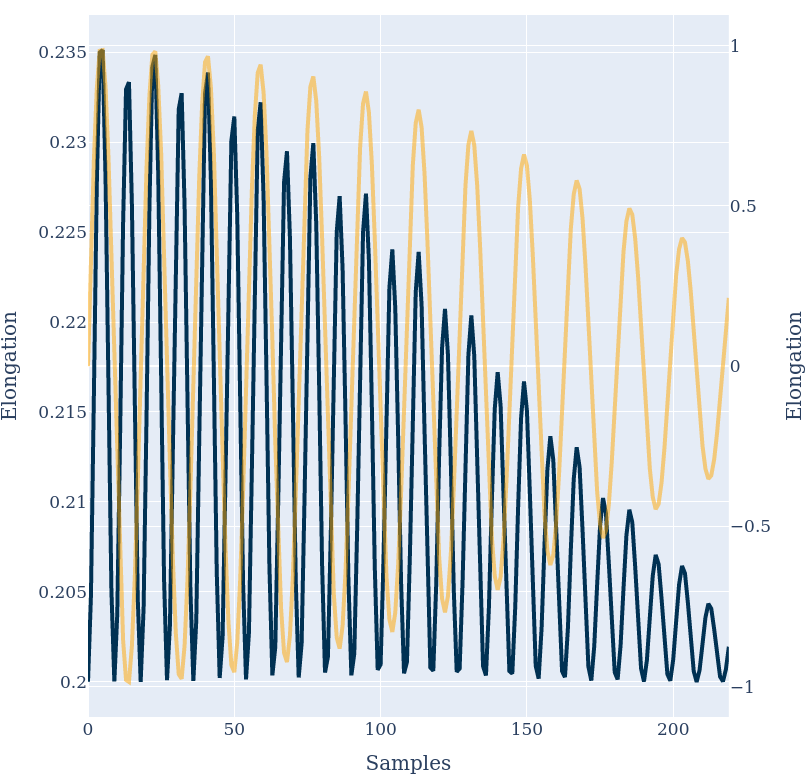}
	\caption{\label{fig:frequency_time_domain} The input signal (orange) and the antenna output signal in the time domain. A frequency doubling similar to Fig. \ref{fig:input_amplitude} can be observed}
\end{figure}

\subsection{Real Mosquitoes}
As it has been shown that mosquito hearing is limited to frequencies below $470$ Hz it has been deduced that mosquitoes respond to the frequency difference of overtones when performing the synchronization. To verify this, we simulated the synchronization experiment of \cite{sciencesongs} 
who observed the synchronization behavior of a male Aedes aegypti when exposed to a simulated female Aedes aegypti's flight tones. We insert the flight tone of a modeled mosquito $m_\text{int}$ superimposed with the flight tone of another mosquito $m_\text{ext}$ (internal vs. external) to deflect the flagellum. The model output is $40$ different time series that represent the stretch and compression of attachment cells at the individual prongs.

\begin{figure}[h!]
	\centering
	\includegraphics[width=3in]{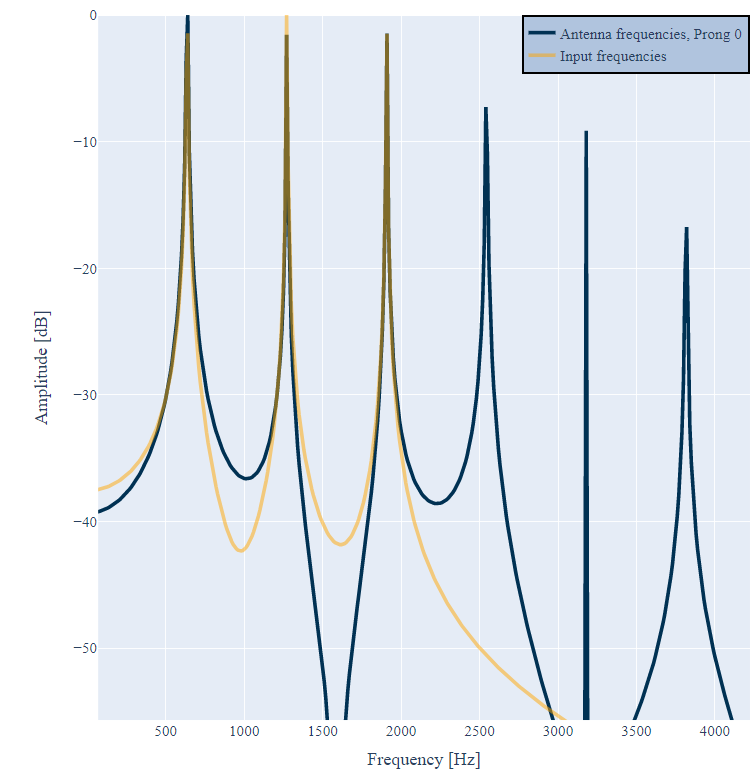}
	\caption{\label{fig:single_int} Spectrum of a artificially created mosquito signal with frequencies of $636$ Hz, $1272$ Hz, $1908$ Hz  and the reaction of its antenna. The orange signal describes the input signal, while the blue signal describes the reaction of the antenna.}
\end{figure}

At first we approximate the natural male mosquito ($m_\text{int}$) wing beat sound with a fundamental frequency of $636$ Hz and its $J=2$ multiples $1272$ Hz and $1908$ Hz. We observed that the amplitudes of overtones of natural mosquitoes decrease. We simulated this property by reducing the amplitude $g_0=1$ of the first overtone to $g_1=0.95$ and the second overtone to $g_2=0.8$
\begin{equation}
m_\text{int} (t) = \sum_{j=1}^{J} g_j \sin (2 \pi ~j~ 636 \text{~Hz~} t) .
\end{equation}
This equation represents the sound input to the antenna in time domain, i.e., $d(t)$, when a mosquito flies. Figure \ref{fig:single_int} illustrates the sound input in frequency domain, i.e., $D(\omega)$, together with the response $R_0(\omega)$. With only the input of the flying mosquito itself, the response only contains the input frequencies plus harmonic distortion. No excitation inside the audible frequency region $<470$ Hz can be observed.

\begin{figure}[h!]
	\centering
	\includegraphics[scale=0.25]{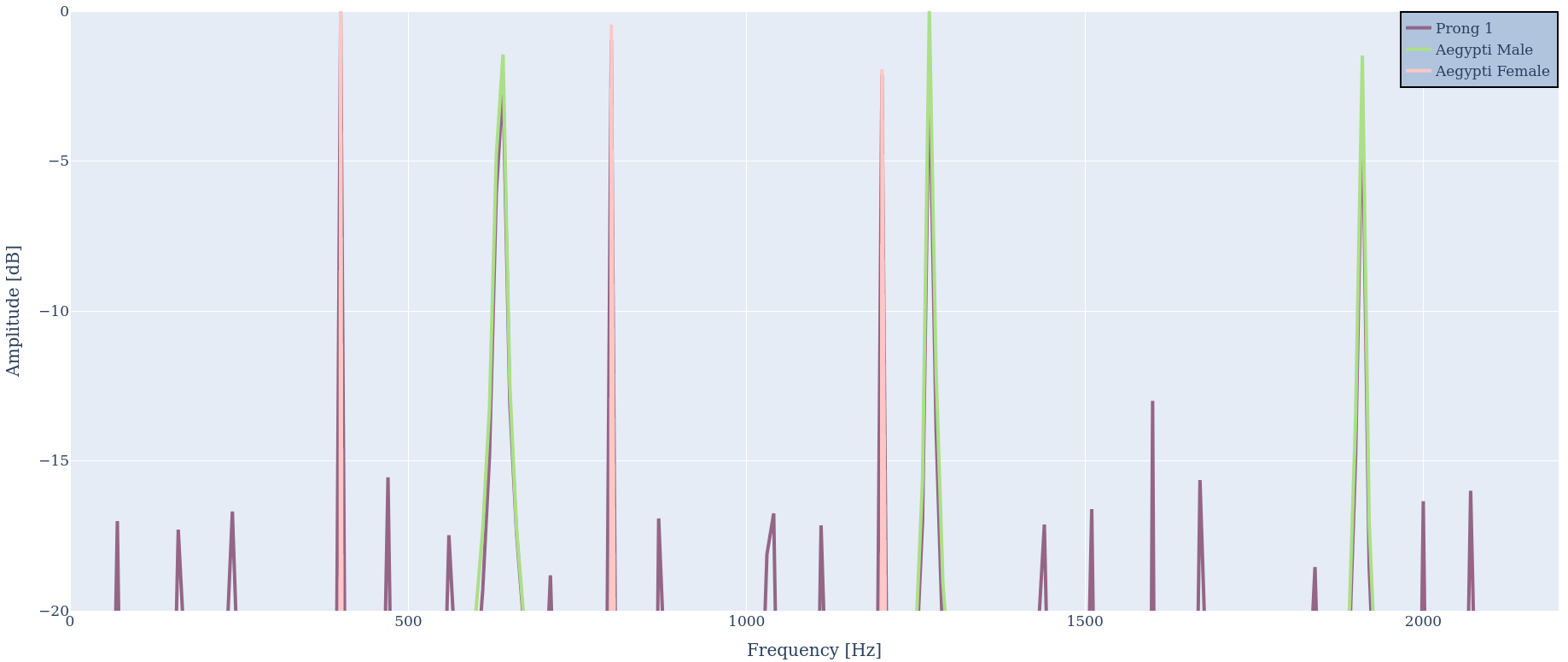}
	\caption{\label{fig:two_mosq_zoomed} Two virtual mosquitoes with fundamental frequencies of $400$ Hz (red) and $636$ Hz (green) serve as an input spectrum $D(\omega)$, which creates the output spectrum $R_1(\omega)$ (purple) that contains the input frequencies plus difference frequencies and harmonic distortion.}
\end{figure}

In the next step we simulate two mosquitoes trying to synchronize with each other. This is illustrated in Fig. \ref{fig:two_mosq_zoomed}. The harmonic spectrum of the male mosquito (green) is represented by the harmonic series of $636$ Hz, $1272$ and $1908$ Hz. The harmonic spectrum of the female mosquito is represented by frequencies of $400$ Hz, $800$ Hz and $1200$ Hz. The male's first harmonic overtone at $1272$ Hz is close to the female's second harmonic overtone at $1200$ Hz, similar to the pure tone example in Fig. \ref{fig:input_amplitude}. The purple signal represents the antenna output $R_1(\omega)$. Again, the antenna output contains all input frequencies. In addition, all difference frequencies can be found, most importantly $1272 \text{~Hz~}-1200\text{~Hz~}=72$ Hz, $800\text{~Hz~}-636\text{~Hz~}=164$ Hz and $636\text{~Hz~}-400\text{~Hz~}=236$ Hz, which lie well inside the audible frequency region of mosquitoes. This is in agreement with criteria 3 and 4.

\begin{figure}[h!]
	\centering
	\includegraphics[scale=0.2]{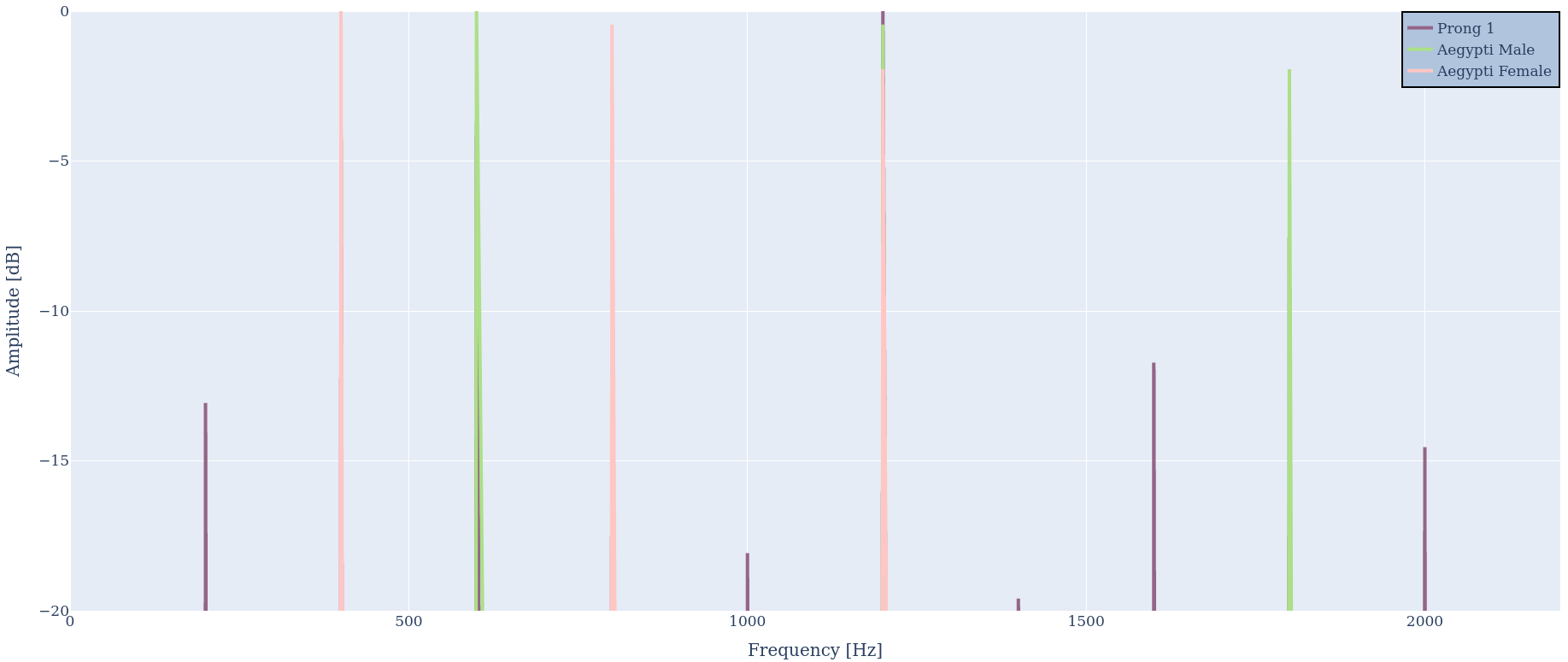}
	\caption{\label{fig:synched_mosquitoes} Perfectly synchronized male (green) and female mosquito (red) frequencies. Again, the input frequencies, harmonic distortion and the difference frequency are visible in the output $R_1(\omega)$ (purple).}
\end{figure}

We reduced the fundamental frequency of the simulated male in steps of $6$ Hz down to the perfect synchronization ratio of $600$ Hz. As expected, the output continues to show the input frequencies, harmonic distortion and all frequency differences. 
The perfectly synchronized mosquito spectra are illustrated in Fig. \ref{fig:synched_mosquitoes}. As can be seen, all low-frequency components vanished, except for the difference between the fundamental frequencies, i.e., $600 \text{~Hz~}-400\text{~Hz~}=200\text{~Hz~}$.

\begin{figure}[h!]
	\centering
	\includegraphics[width=3in]{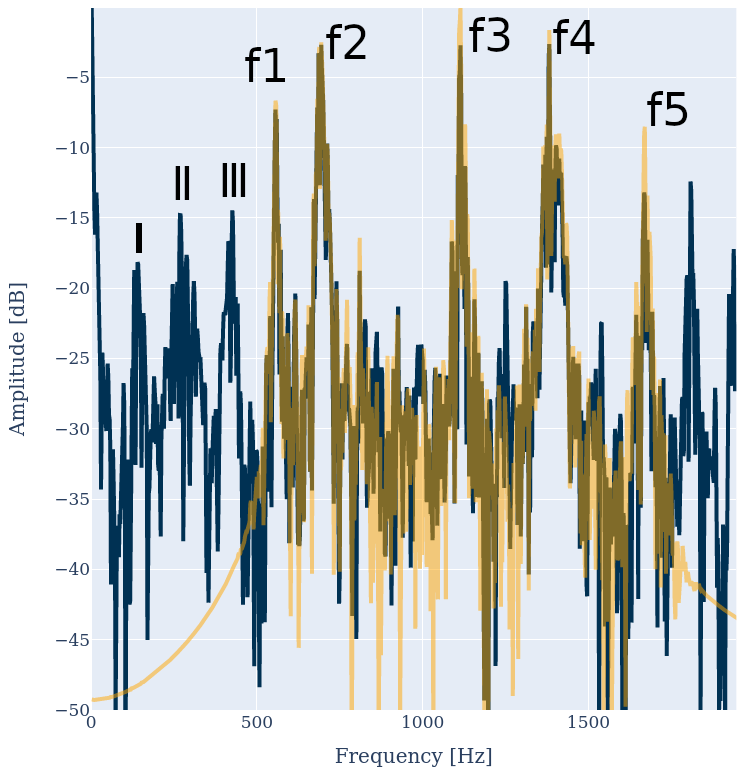}
	\caption{\label{fig:delta_f0} Spectrum of two synchronizing mosquitoes (orange) and response at prong 0 (blue). The female frequencies are f1=$555$ Hz, f3=$1112$ Hz and f5=$1665$ Hz, the male frequencies are f2=$688$ Hz and f4=$1383$ Hz. Their difference frequencies \RomanNumeralCaps{1}=f2-f1=$133$ Hz, the broad peak \RomanNumeralCaps{2}=f4-f3=271} Hz and f5-f4=$282$ Hz and \RomanNumeralCaps{3}=f3-f2=$424$ Hz are clearly visible.
\end{figure}

As a next step, we analyzed a recording of real mosquitoes during their synchronization process. The recorded spectrum and the response at prong $0$ are illustrated in Fig. \ref{fig:delta_f0}. We used a bandpass filter between $550$ and $1700$ Hz to eliminate the noise outside the relevant frequency region of mosquito mating. As can be seen, the signal-to-noise ratio lies in the order of $25$ dB. The model output exhibits the same characteristics as observed with artificial mosquito spectra: Over a wide frequency range, the model output resembles the model input. In addition, the output contains harmonic distortion, i.e., high-frequency peaks on the right of the original input spectrum. Most importantly, the antenna response contains the difference frequencies. The broad low-frequency peaks \RomanNumeralCaps{1}, \RomanNumeralCaps{2} and \RomanNumeralCaps{3} correspond to the difference frequencies \RomanNumeralCaps{1}=f2-f1, the proximate frequencies \RomanNumeralCaps{2}=f4-f3 and f5-f4, and \RomanNumeralCaps{3}=f2-f2. 

Note that the peaks at f1 to f5 are no distinct peaks, but rather double or even quadruple peaks, as both mosquitoes altered their wing beat rate during the recording. The difference frequencies of such double peaks are also represented near the lower end of the output spectrum, like the peaks near $0$ Hz and $25$ Hz. These may provide valuable synchronization information for the mosquito.

\subsection{Localization Cues}
As mosquitoes are very small compared to the wavelengths that their wing beats produce, we model them as point source \citep{complexp}. Here, the inverse distance law is valid for sound pressure, i.e., the amplitude is proportional to the inverse of the distance $r$, like
\begin{equation}
    \hat{A} \propto \frac{1}{r} \ .
    \label{eq:inverse}
\end{equation}
We simulated a male mosquito ($m_\text{ext}$) with three partials as before, and a female mosquito ($m_\text{int}$) with the different partials. This yields six input frequencies, three of which are produced by the modeled mosquito, three, by the other mosquito. Following \cite{complexp} the input spectra at one antenna are calculated by superposition of the propagated spectrum of the female mosquito with the spectrum of the male mosquito like

\begin{equation}
\label{eq:1r}
D(\omega)=M_{\text{ext}} (\omega)\frac{e^{i k r}}{r}+M_{\text{int}} (\omega) \ ,
\end{equation}
where $e\approx 2.718$ is Euler's number, $i$ is the imaginary unit defined as $i^2=-1$ and $r$ is the Euclidean distance between sound source and each antenna.

We modeled two antennae $0.125$ mm apart from each other, which is a typical antenna distance \citep{active}. Then, we put the external mosquito at three different angles of $30^\circ$, $60^\circ$ and $90^\circ$ at two different distances, namely $1.5$ cm and $3$ cm away from the center of the antennae. The results of this simulation are summarized in Table \ref{tab:ild}.

The amplitudes of the input spectra $\hat{A}_{f_\text{in}}$ as well as the Inter antennal Amplitude Differences (IAD) of the input spectra $\text{IAD}_{f_{\text{in}}}$, are determined by Eq. \ref{eq:1r}. When doubling the distance of the sound source, the amplitude reduces by approximately $6$ dB. In motion, this is a distance cue that is also utilized by humans and other animals \cite[chap. 4]{book}. The IAD of the input frequencies depends on distance and angle. As can be seen in the Table, it may lie in a range below $0.072$ dB. In humans and other animals, Inter aural Level Differences (ILD) inform about the direction of a sound source \cite[chap. 4]{book}.

So the relationship between $\hat{A}$ and distance as well as the relationship between IAD and angle should provide mosquitoes with localization cues. However, in order to be utilizable by a mosquito, these relationships need to take place at frequencies below $470$ Hz. Our virtual male mosquito does not produce a single frequency inside the audible hearing range of the modeled female mosquito. But as could be seen already in Fig. \ref{fig:two_mosq_zoomed}, the superposition of the female and male flight tones produces three difference frequencies $f_\text{out}$, namely $72$ Hz, $164$ Hz and $236$ Hz, in addition to the three frequencies $f_\text{in}$ from the male mosquito $m_\text{ext}$. In Table \ref{tab:ild} we provide the $\hat{A}$ and IAD of $f_\text{out}$ together with those of $f_\text{in}$. It can be seen that the localization cues are almost equal, i.e., they are perfectly transferred to the difference frequencies that lie below $470$ Hz. This fulfills criteria 5, i.e., the antenna produces audible sound source localization cues due to the nonlinearities that produce difference frequencies.



\begin{table}[]
    \centering
    \begin{tabular}{| c || c | c || c | c |}
      \hline
      Angle & \multicolumn{2}{|c|}{$1.5$ cm} & \multicolumn{2}{|c|}{$3$ cm} \\
      \hline
      &$\hat{A}_{f_\text{in}}\pm \text{IAD}_{f_{\text{in}}}$&$\hat{A}_{f_\text{out}}\pm \text{IAD}_{f_{\text{out}}}$&$\hat{A}_{f_\text{in}}\pm \text{IAD}_{f_{\text{in}}}$&$\hat{A}_{f_\text{out}}\pm \text{IAD}_{f_{\text{out}}}$\\
      \hline
      $\pm 30^\circ$ & $-0.02 \pm 0.036$ dB &$-0.02 \pm 0.036$ dB &$-5.98 \pm 0.018$ dB & $-6.04\pm 0.031$ dB \\ 
      \hline
      $\pm 60^\circ$ & $-0.01 \pm 0.062$ dB& $-0.00 \pm 0.062$ dB & $-5.89 \pm 0.031$ dB& $-6.03 \pm 0.031$ dB\\ 
      \hline
      $\pm 90^\circ$ & $0.00 \pm 0.072$ dB &  $0.00 \pm 0.072$ dB & $-5.97 \pm 0.036$ dB & $-6.03 \pm 0.036$ dB\\
      \hline
    \end{tabular}
    \caption{Strongest inter antennal amplitude difference per angle calculated for the three difference tones between two flying mosquitoes. Especially near the most sensitive frequency region the difference offers a robust localization cue.}
    \label{tab:ild}
\end{table}

\section{Discussion}
Our model is based on the geometry of the mosquito antenna morphology. The fact that this model is able to produce all $5$ observations that have been made in real mosquitoes provides evidence that many of the complicated nonlinearities can be attributed already to the periphery of mosquito hearing, i.e., the antenna base, rather than to purely neural processing. This has many implications. Most importantly, the model 
allows us to reflect findings from a new perspective and to approach answers to open questions on the basis of antenna base morphology, rather than on the basis of flagellum mechanics or neural efferents.

For example \cite{sciencesongs} carried out synchronization experiments between real mosquito pairs and the synchronization of a mosquito to artificial sounds played through an ear bud speaker and measured neural responses in the JO of a male mosquito. They conclude that the upper limit of hearing in mosquitoes is $2000$ Hz and not the widely accepted $470$ Hz, because responses were even measurable when the artificial sound contained only frequencies above $800$ Hz. Naturally, the flight tone of the male mosquito, whose neural response to stimuli has been measured, contains no frequencies below the $470$ Hz either. However, they mention that they were able to detect neural response to high-frequency inputs, when setting their high-pass filter down to $1$ or less Hz, but not when choosing $100$ Hz, i.e., they detected low-frequency neural activity as a response to a superposition of high-frequency sounds. Our model suggests that the auditory neurons in mosquitoes do not directly respond to such high frequencies, but rather to the frequency differences produced by the antenna as shown in our model. Only through the interplay of the male mosquito flight tone and the artificial signal, the antenna produces motions with frequency components below $100$ Hz that excite the neurons. Consequently, the neural activity is not a response to high-frequency input, but to the frequency difference between multiple high-frequency inputs that create a low-frequency response ($<100$ Hz) already in the antenna. The neuronal low-frequency response simply represents the low-frequency motion that is mechanically induced by the nonlinear transform from deflection to rotation inside the antenna base. Mosquitoes detect the frequency difference and try to minimize it, to achieve a synchronization.

\cite{Lapshin3927} found anti-phase responses in electrical measurements inside the JO and already speculated that the mechanical motion of the antenna may cause them. Our model supports this speculation and provides evidence that these anti-phase responses result from the geometry of the antenna base morphology, namely the curved septa.

\cite{invivowindmill} wrote that "The presence of a strong frequency doubled component in an otherwise undistorted signal reinforces the hypothesis that, rather than a by-product of some unknown mechanical property, this effect is strongly correlated with the active neuromechanical amplification produced by the Johnston’s organ’s neurons." Even though active neuromechanical amplification is likely to happen, our model sheds light on the ``unknown mechanical property'', indicating that the transfer from deflection to rotation in the antenna base may cause the frequency doubling at some prong locations.

\begin{figure}[h!]
	\centering
	\includegraphics[width=3in]{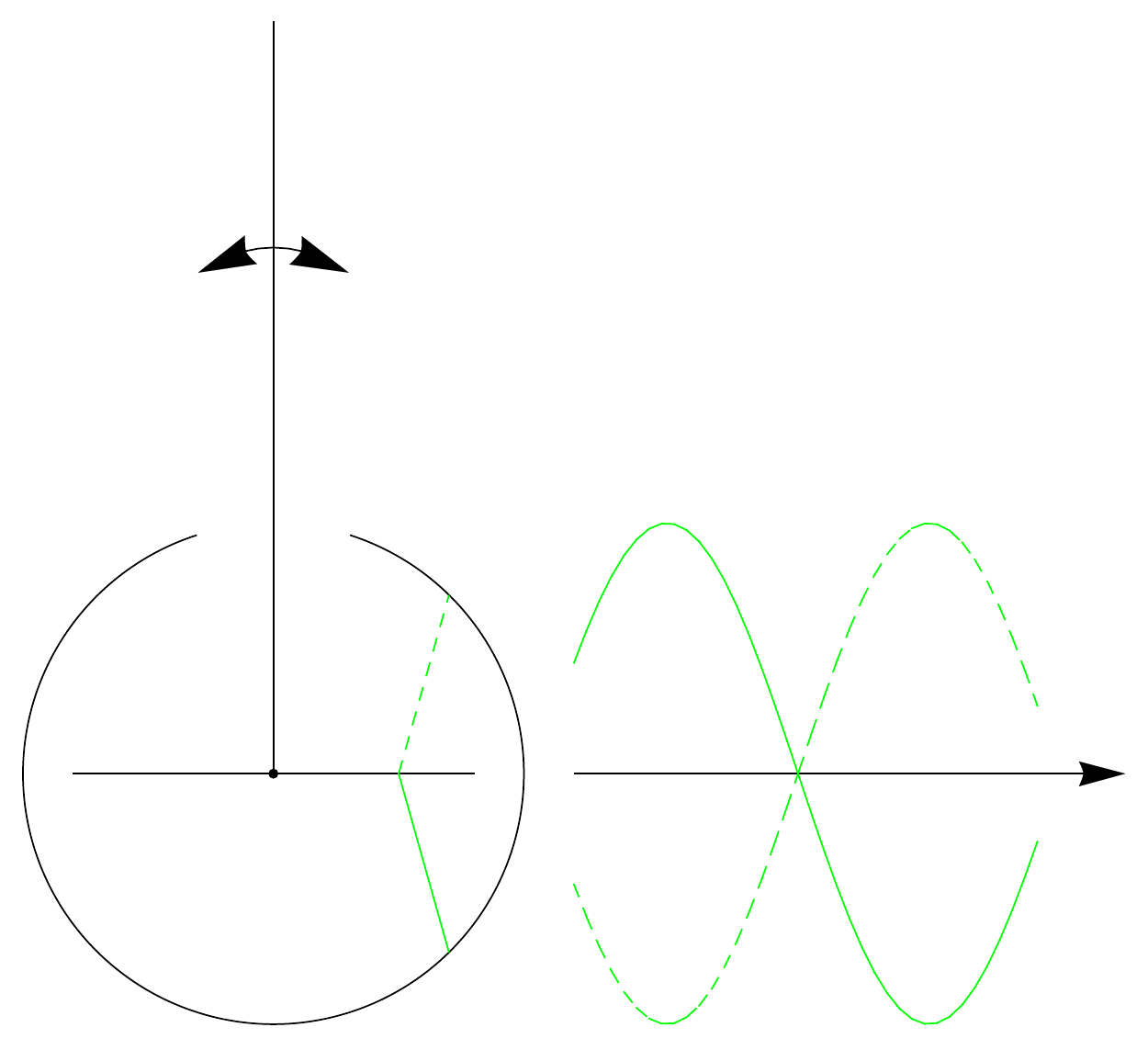}
	\caption{\label{fig:disto} Antenna model by \cite{distortion}: symmetric attachment cells create anti-phase responses.}
\end{figure}

The antenna model from \cite{distortion} is illustrated in Fig. \ref{fig:disto}. The model simplifies the basilar plate and septa as a line rather than a curve. Here, the authors connect the septa to the JO in the pedicel by pairs of attachment cells above and below the septa. This symmetry creates anti-phase responses, as the attachment cells above the basilar plate are compressed when the attachment cells below the basilar plate are stretched and vice versa. However, the actual mosquito antenna does not exhibit this symmetry of attachment cells. In contrast, the geometry in our model is in accordance with the actual morphology of the mosquito antenna but still produces anti-phase responses.

The model in \cite{mechan} looks very similar but lacks the pairwise attachment cells. The authors had to add a neural model to produce nonlinearities. Our model produces such nonlinearities already without including neural afferents or efferents. It could therefore be a more realistic pre-processor for neural mosquito hearing models.

\cite{roth1948} assumed that the long fibrillae in male mosquito antennas resonate with female wing beat frequencies, enabling them to detect females.   This assumption seems plausible, as male mosquitoes have a much bushier antenna than female mosquitoes. However, this assumption does not sufficiently explain how female mosquitoes are able to synchronize to male mosquitoes at inaudibly high frequencies. Our model does not support or contradict the assumption of resonating fibrillae, but it provides evidence that even in the absence of fibrillae, the antenna clearly indicates the presence and even the location of a mosquito through nonlinear interplay of two flight sounds. This means the model explains –– in a reasonable, morphologically plausible way –– how both male and female mosquitoes are able to detect the presence and location of one another.

From acoustical and mechanical considerations \cite{tischner} derived that a direction-dependent distortion factor enables mosquitoes to localize sound sources. This hypothesis was supported by electrode measurements inside the pedicel during magnetic antenna stimulation \citep{richtung}. Our model exhibits harmonic distortions whose intensity depends on distance rather than on direction. However, as harmonic distortion only produces frequencies outside the hearing range of mosquitoes, it is unlikely that mosquitoes can actually make use of such cues. Instead, our model indicates that the nonlinear transform from deflection to rotation inside the antenna base produces difference frequencies in the most-sensitive frequency region of mosquitoes. Furthermore, inter antennal amplitude differences have been observed. These are angle dependent and provide the mosquito with a localization cue for source direction, just as inter aural level differences in human hearing. The existence of inter antennal cues in our antenna base model may explain why mosquitoes have an antenna pair and not just one single antenna. Consequently, our model suggests that localization may be realized through a combination of mono antennal and biantennal cues, just as the monaural and binaural cues in human sound source localization.


\cite{simuldoubl} measured neural responses to single frequencies using glass electrodes in the JO. They compared the threshold for passive mosquitoes with the threshold during simulated mosquito flight. They observed that the threshold of neural response typically dropped by $8$ dB within the frequency region from $80$ to $120$ Hz for flying mosquitoes compared to resting mosquitoes. One exception was a frequency around $40$ Hz. Here, the threshold dropped on average by $26$ dB (maximum by $56$ dB (!)). Our model provides evidence that this low-frequency component is prominently created through the nonlinear combination of internal and external mosquito wingbeat sounds inside the antenna base during courtship. This explains how this amazing amplification can be achieved through mechanical pre-processing. Furthermore, our model yields a possible explanation why this frequency is amplified so much: It appears that mosquitoes synchronize by minimizing the low-frequency component. Consequently, the beat frequency that results from proximate frequencies near the synchronization frequency (roughly in the order $\leq 72$ Hz) has to be amplified much stronger than the other beat frequencies, like the difference between the fundamental frequencies, e.g., $636$ Hz$-400 \text{~Hz~} = 236$ Hz.

Furthermore, \cite{simuldoubl} could show that the acoustic sensitivity of both female and male mosquitoes increased considerably during simulated flight, even though females exhibit much fewer fibrillae and auditory neurons. Our model provides evidence that a lot of the nonlinear processing during flight results from the transformation of deflection to rotation inside the antenna base, rather than at the level of the fibrillar or neural processing: In contrast to the sexually dimorphic flagellum, the antenna base is morphological similar in female and male mosquitoes.


\cite{julia} modeled the mechanics of two mosquito flagellum shafts as rigid beams with one free and one clamped end. She could show that magnitude and phase relationship between both antenna deflections provides a strong and unique directional cue. She derived this from simulations with an inaudibly high frequency of an external male, in absence of an internal female sound. However, our model provides evidence that the deflections stay represented in the low-frequency domain when adding an internal female sound. The combination of her physical flagellum shaft model and our antenna base model could provide stronger and more plausible sound source localization cues than each of the individual models.

\cite{Lapshin3927} observed that many auditory neurons in the JO are tuned to frequencies between $190$ to $270$ Hz, which corresponds to the $f_0$ difference of male and female mosquitoes. Our model provides evidence that this difference frequency is produced inside the antenna base through the transfer from deflection to rotation. Moreover, other difference frequencies fall into this region, too, especially during the act of synchronization. Our localization experiment adds another possible explanation for the high number of neurons that are tuned to this frequency region: inter antennal amplitude differences are most prominent in this frequency region, so it is likely that not only neurons responsible for frequency detection but also for source localization are tuned to these frequencies.

\cite{simuldoubl} found the frequency deviation and twice the frequency deviation neurally represented during simulated flight experiments. Our model suggests that these deviations are already produced mechanically in the antenna base and only passively represented or actively amplified through auditory neurons.

In contrast to many mosquito species --- like Aedes aegpty and Aedes albopictus (dengue vectors), Anopheles dirus and Anopheles minimus (malaria vectors) and Culex quinquefasciatus (non-disease vector) \citep{lyn} --- the fundamental frequency of male and female Tachina brevipalpis is similar. Still, \cite{f0sync} observed that mosquitoes of both sexes would alter their wing beat frequency to synchronize with a pure tone as long as its frequency deviated less that $60$ Hz from their own fundamental frequency. They conclude that mosquitoes can detect many frequency deviations, i.e., beat frequencies between many frequency pairs, to synchronize. Our model supports this assumption, as not only the beat frequency of overtones near the typical synchronization frequency of $1200$ Hz is represented by the antenna response, but also the deviation between fundamental frequencies and between proximate partials. It appears that mosquitoes' strategy during synchronization is to minimize the lowest frequency component, which appears to be difference between the frequencies that lie closest to the favored synchronization frequency.

Given these results, we suggest to use an antenna model with a curved septa as a biologically-inspired pre-processor rather than using the raw audio as an input to neural models of mosquito hearing. Furthermore, we suggest not to underestimate the importance of the antenna base in addition to investigations that concentrate on flagellum mechanics and auditory efferents. Our model provides evidence that the mosquito antenna base delivers reliable cues that enable mosquitoes to detect the presence of other mosquitoes, to localize them, and to synchronize by means of minimizing the lowest frequency represented in the antenna response.

As acoustics seem to be \textcolor{red}{an important} sexual stimulant in mosquitoes, understanding their hearing system may be helpful in the future to support the fight against vector-borne diseases, like malaria, dengue fever and yellow fever by means of:
\begin{itemize}
    \item automatic acoustic identification of  mating swarm sites
    \item acoustically dispelling mosquitoes
    \item species- and sex-specific luring into traps for
    \begin{itemize}
        \item manual or 
        \item automatic mosquito monitoring (recognition and counting of vector species)
        \item SIT and IIT preparation
    \end{itemize}
\end{itemize}

\section{Conclusion}
In this paper we introduced a biologically inspired mosquito antenna model. The model exhibits many of the characteristics that have been observed through neural and behavioral studies of real mosquitoes: The model produces harmonic distortions whose number and intensity depend on the intensity of the input signal and the location of the prong. Some prongs exhibit an anti-phase response, which has been observed in neural measurements of real mosquitoes. The model produces the difference frequencies between the fundamental frequency of the male and female mosquito as well as the frequency difference between overtones. These frequency differences lie well inside the hearing range of a mosquito, even if the input signal lies outside their hearing range. The intensity of the difference frequencies is distance-dependent, providing a source distance cue. When modeling two antennae, the model produces inter-antennal level differences at the beat frequency, providing localization cues comparable to the inter-aural level differences of humans and other animals with paired ears.

The fact that our geometric model meets all of the $5$ above-mentioned criteria that have been observed in real mosquitoes is evidence that the geometry of the antenna morphology plays a crucial role in mosquito hearing. Already the periphery produces many of the observations that had been attributed to neural activity and efferents (see e.g. \citep{Jackson16734}, \citep{active} and \citep{current}). These nonlinearities enable mosquitoes to detect, localize, and synchronize to one another, i.e., to carry out their courtship behavior which is mandatory for mating. The model made the important role of the antenna base evident, which is underestimated, or at least underrepresented in the study of mosquito hearing and courtship behavior.

The current state of the model neglects physical properties which may be included as filters in the future. Neurally, the stretch and compression of the attachment cells, as calculated in our model, is neurally encoded and processed further inside the mosquito's auditory system. This processing has to be implemented in the future to gain an even deeper understanding of mosquito hearing and courtship behavior.
\section*{Conflict of Interest Statement}
The authors declare that the research was conducted in the absence of any commercial or financial relationships that could be construed as a potential conflict of interest.

\section*{Author Contributions}
Tim Ziemer conceptualized and implemented the original mosquito antenna model and started to evaluate it with artificial sounds. Fabian Wetjen and Alexander Herbst reimplemented the model, evaluated it further, with artificial mosquitoes and with real mosquito recordings. All authors contributed equally to the literature research, manuscript preparation and audio analysis.

\section*{Acknowledgments}
We thank Thomas Barkowsky and the \href{http://www.mobile4d.capacitylab.org}{Mobile4D-team} for fruitful discussions and continuous feedback. We also thank Myat Su Yin, Peter Haddawy and their team at \href{https://miru.ict.mahidol.ac.th}{MIRU} for continuous exchange of ideas and mosquito recordings. Lastly, we thank Paul Barmsen from \href{https://tinydrops.de/}{tiny drops}, who introduced us to his mosquito control network and provided us with additional mosquito recordings.


\section*{Data Availability Statement}
The model of the antenna base can be found in the [MOSQUITO ANTENNA REPOSITORY] [\url{https://github.com/HughIdiyit/mosquito_antenna}].

\bibliographystyle{frontiersinSCNS_ENG_HUMS} 
\bibliography{frontiers}

\begin{thebibliography}{47}
\providecommand{\natexlab}[1]{#1}
\expandafter\ifx\csname urlstyle\endcsname\relax
  \providecommand{\doi}[1]{doi:\discretionary{}{}{}#1}\else
  \providecommand{\doi}{doi:\discretionary{}{}{}\begingroup
  \urlstyle{rm}\Url}\fi
\providecommand{\selectlanguage}[1]{\relax}
\providecommand{\bibAnnoteFile}[1]{%
  \IfFileExists{#1}{\begin{quotation}\noindent\textsc{Key:} #1\\
  \textsc{Annotation:}\ \input{#1}\end{quotation}}{}}
\providecommand{\bibAnnote}[2]{%
  \begin{quotation}\noindent\textsc{Key:} #1\\
  \textsc{Annotation:}\ #2\end{quotation}}

\bibitem[{Albert and Kozlov(2016)}]{comparative}
Albert, J.~T. and Kozlov, A.~S. (2016).
\newblock Comparative aspects of hearing in vertebrates and insects with
  antennal ears.
\newblock \emph{Current Biology} 26, R1050--R1061.
\newblock \doi{10.1016/j.cub.2016.09.017}
\bibAnnoteFile{comparative}

\bibitem[{Alphey et~al.(2010)Alphey, Benedict, Bellini, Clark, Dame, Service
  et~al.}]{sit}
Alphey, L., Benedict, M., Bellini, R., Clark, G.~G., Dame, D.~A., Service,
  M.~W., et~al. (2010).
\newblock Sterile-insect methods for control of mosquito-borne diseases: An
  analysis.
\newblock \emph{Vector-Borne and Zoonotic Diseases} 10, 295--311.
\newblock \doi{10.1089/vbz.2009.0014}.
\newblock PMID: 19725763
\bibAnnoteFile{sit}

\bibitem[{Andrés et~al.(2016)Andrés, Seifert, Spalthoff, Warren, Weiss,
  Giraldo et~al.}]{current}
Andrés, M., Seifert, M., Spalthoff, C., Warren, B., Weiss, L., Giraldo, D.,
  et~al. (2016).
\newblock Auditory efferent system modulates mosquito hearing.
\newblock \emph{Current Biology} 26, 2028--2036.
\newblock \doi{10.1016/j.cub.2016.05.077}
\bibAnnoteFile{current}

\bibitem[{Arthur et~al.(2010)Arthur, Wyttenbach, Harrington, and
  Hoy}]{distortion}
Arthur, B.~J., Wyttenbach, R.~A., Harrington, L.~C., and Hoy, R.~R. (2010).
\newblock Neural responses to one- and two-tone stimuli in the hearing organ of
  the dengue vector mosquito.
\newblock \emph{J Exp Biol} 213, 1376--1385.
\newblock \doi{10.1242/jeb.033357}
\bibAnnoteFile{distortion}

\bibitem[{Avitabile et~al.(2010)Avitabile, Homer, Champneys, Jackson, and
  Robert}]{mechan}
Avitabile, D., Homer, M., Champneys, A.~R., Jackson, J.~C., and Robert, D.
  (2010).
\newblock Mathematical modelling of the active hearing process in mosquitoes.
\newblock \emph{Journal of The Royal Society Interface} 7, 105--122.
\newblock \doi{10.1098/rsif.2009.0091}
\bibAnnoteFile{mechan}

\bibitem[{Becker et~al.(2010)Becker, Petric, Zgomba, Boase, Dahl, Madon
  et~al.}]{buch}
Becker, N., Petric, D., Zgomba, M., Boase, C., Dahl, C., Madon, M., et~al.
  (2010).
\newblock \emph{Mosquitoes and Their Control} (Berlin Heidelberg: Springer),
  second edn.
\newblock \doi{10.1007/978-3-540-92874-4}
\bibAnnoteFile{buch}

\bibitem[{Berkhausen(2021)}]{thesis}
Berkhausen, B. (2021).
\newblock \emph{Modeling of a mosquito antenna in a soundfield in Python}.
\newblock Bachelor's thesis, University of Hamburg, Bremen
\bibAnnoteFile{thesis}

\bibitem[{Boo and Richards(1975)}]{structure}
Boo, K.~S. and Richards, A. (1975).
\newblock Fine structure of the scolopidia in the johnston's organ of male
  aedes aegypti (l.) (diptera: Culicidae).
\newblock \emph{International Journal of Insect Morphology and Embryology} 4,
  549--566.
\newblock \doi{https://doi.org/10.1016/0020-7322(75)90031-8}
\bibAnnoteFile{structure}

\bibitem[{Cator et~al.(2009)Cator, Arthur, Harrington, and Hoy}]{sciencesongs}
Cator, L.~J., Arthur, B.~J., Harrington, L.~C., and Hoy, R.~R. (2009).
\newblock Harmonic convergence in the love songs of the dengue vector mosquito.
\newblock \emph{Science} 323, 1077--1079.
\newblock \doi{10.1126/science.1166541}
\bibAnnoteFile{sciencesongs}

\bibitem[{Chen et~al.(2014)Chen, Why, Batista, Mafra-Neto, and Keogh}]{chen}
Chen, Y., Why, A., Batista, G., Mafra-Neto, A., and Keogh, E. (2014).
\newblock Flying insect classification with inexpensive sensors.
\newblock \emph{Journal of Insect Behavior} 27, 657--677.
\newblock \doi{10.1007/s10905-014-9454-4}
\bibAnnoteFile{chen}

\bibitem[{Davis and Sokolove(1976)}]{Davis1976}
Davis, E.~E. and Sokolove, P.~G. (1976).
\newblock Lactic acid-sensitive receptors on the antennae of the mosquito,
  aedes aegypti.
\newblock \emph{Journal of comparative physiology} 105, 43--54.
\newblock \doi{10.1007/BF01380052}
\bibAnnoteFile{Davis1976}

\bibitem[{Diabate and Tripet(2015)}]{mating}
Diabate, A. and Tripet, F. (2015).
\newblock Targeting male mosquito mating behaviour for malaria control.
\newblock \emph{Parasites \& vectors} 8, paper number 347.
\newblock \doi{10.1186/s13071-015-0961-8}
\bibAnnoteFile{mating}

\bibitem[{Gibson et~al.(2010)Gibson, Warren, and Russell}]{f0sync}
Gibson, G., Warren, B., and Russell, I.~J. (2010).
\newblock Humming in tune: Sex and species recognition by mosquitoes on the
  wing.
\newblock \emph{Journal of the Association for Research in Otolaryngology} 11,
  527--540.
\newblock \doi{10.1007/s10162-010-0243-2}
\bibAnnoteFile{f0sync}

\bibitem[{Göpfert et~al.(1999)Göpfert, Briegel, and Robert}]{besthearing}
Göpfert, M.~C., Briegel, H., and Robert, D. (1999).
\newblock Mosquito hearing: Sound-induced antennal vibrations in male and
  female aedes aegypti.
\newblock \emph{J. Exp. Biol.} 202, 2727--2738
\bibAnnoteFile{besthearing}

\bibitem[{Göpfert and Robert(2001)}]{active}
Göpfert, M.~C. and Robert, D. (2001).
\newblock Active auditory mechanics in mosquitoes.
\newblock \emph{Proc. R. Soc. Lond. B.} 268, 333--339.
\newblock \doi{10.1098/rspb.2000.1376}
\bibAnnoteFile{active}

\bibitem[{Jackson and Robert(2006)}]{Jackson16734}
Jackson, J.~C. and Robert, D. (2006).
\newblock Nonlinear auditory mechanism enhances female sounds for male
  mosquitoes.
\newblock \emph{Proceedings of the National Academy of Sciences} 103,
  16734--16739.
\newblock \doi{10.1073/pnas.0606319103}
\bibAnnoteFile{Jackson16734}

\bibitem[{Keppler(1958)}]{richtung}
Keppler, E. (1958).
\newblock Über das richtungshören von stechmücken.
\newblock \emph{Z. Naturforsch.} 13, 280--284
\bibAnnoteFile{richtung}

\bibitem[{Koch(2021)}]{julia}
Koch, J. (2021).
\newblock A bio-inspired approach to explore the workings of the mosquito's
  hearing organ [bachelor's thesis].
\newblock \emph{University of Bremen}
\bibAnnoteFile{julia}

\bibitem[{Lapshin(2012)}]{simuldoubl}
Lapshin, D.~N. (2012).
\newblock The auditory system of blood-sucking mosquito females (diptera,
  culicidae): Acoustic perception during flight simulation.
\newblock \emph{Entomological Review} \doi{10.1134/S0013873813020012}
\bibAnnoteFile{simuldoubl}

\bibitem[{Lapshin and Vorontsov(2017)}]{Lapshin3927}
Lapshin, D.~N. and Vorontsov, D.~D. (2017).
\newblock Frequency organization of the johnston{\textquoteright}s organ in
  male mosquitoes (diptera, culicidae).
\newblock \emph{Journal of Experimental Biology} 220, 3927--3938.
\newblock \doi{10.1242/jeb.152017}
\bibAnnoteFile{Lapshin3927}

\bibitem[{Matthews and Matthews(2010)}]{insect}
Matthews, R.~W. and Matthews, J.~R. (2010).
\newblock \emph{Insect Behavior} (Dordrecht: Springer), 2nd edn.
\newblock \doi{10.1007/978-90-481-2389-6}
\bibAnnoteFile{insect}

\bibitem[{Menda et~al.(2019)Menda, Nitzany, Shamble, Wells, Harrington, Miles
  et~al.}]{farfield}
Menda, G., Nitzany, E.~I., Shamble, P.~S., Wells, A., Harrington, L.~C., Miles,
  R.~N., et~al. (2019).
\newblock The long and short of hearing in the mosquito aedes aegypti.
\newblock \emph{Current Biology} 29, 709 -- 714.e4.
\newblock \doi{https://doi.org/10.1016/j.cub.2019.01.026}
\bibAnnoteFile{farfield}

\bibitem[{Mhatre(2015)}]{review}
Mhatre, N. (2015).
\newblock Active amplification in insect ears: mechanics, models and molecules.
\newblock \emph{J Comp Physiol A} 201, 19--37.
\newblock \doi{10.1007/s00359-014-0969-0}
\bibAnnoteFile{review}

\bibitem[{Nijhout(1977)}]{erect}
Nijhout, H.~F. (1977).
\newblock Control of antennal hair erection in male mosquitoes.
\newblock \emph{Biological Bulletin} 153, 591--603
\bibAnnoteFile{erect}

\bibitem[{Offenhauser and Kahn(1949)}]{diseasesound}
Offenhauser, W.~H. and Kahn, M.~C. (1949).
\newblock The sounds of disease‐carrying mosquitoes.
\newblock \emph{The Journal of the Acoustical Society of America} 21, 462--463.
\newblock \doi{10.1121/1.1917085}
\bibAnnoteFile{diseasesound}

\bibitem[{Roth(1948)}]{roth1948}
Roth, L.~M. (1948).
\newblock A study of mosquito behavior. an experimental laboratory study of the
  sexual behavior of aedes aegypti (linnaeus).
\newblock \emph{The American Midland Naturalist} 40, 265--352.
\newblock \doi{10.2307/2421604}
\bibAnnoteFile{roth1948}

\bibitem[{Saltin et~al.(2019)Saltin, Matsumura, Reid, Windmill, Gorb, and
  Jackson}]{stiffness}
Saltin, B.~D., Matsumura, Y., Reid, A., Windmill, J.~F., Gorb, S.~N., and
  Jackson, J.~C. (2019).
\newblock Material stiffness variation in mosquito antennae.
\newblock \emph{Journal of The Royal Society Interface} 16, 20190049.
\newblock \doi{10.1098/rsif.2019.0049}
\bibAnnoteFile{stiffness}

\bibitem[{Schwartzkopff(1962)}]{schwartzkopff}
Schwartzkopff, J. (1962).
\newblock Die akustische lokalisation bei tieren.
\newblock In \emph{Ergebnisse der Biologie}, eds. H.~Autrum, E.~B{\"u}nning,
  K.~v.~Frisch, E.~Hadorn, A.~K{\"u}hn, E.~Mayr, A.~Pirson, J.~Straub,
  H.~Stubbe, W.~Weidel, and H.~Autrum (Berlin, Heidelberg: Springer Berlin
  Heidelberg), 136--176
\bibAnnoteFile{schwartzkopff}

\bibitem[{Senevirathna et~al.(2020)Senevirathna, Udayanga, Ganehiarachchi,
  Hapugoda, Ranathunge, and Gunawardene}]{rearing}
Senevirathna, U., Udayanga, L., Ganehiarachchi, G. A. S.~M., Hapugoda, M.,
  Ranathunge, T., and Gunawardene, N.~S. (2020).
\newblock Development of an alternative low-cost larval diet for mass rearing
  of aedes aegypti mosquitoes.
\newblock \emph{Biomed Res Int} , Paper number:
  1053818\doi{10.1155/2020/1053818}
\bibAnnoteFile{rearing}

\bibitem[{Su et~al.(2018)Su, Andres, Boyd-Gibbins, Somers, and
  Albert}]{modelseries}
Su, M.~P., Andres, M., Boyd-Gibbins, N., Somers, J., and Albert, J.~T. (2018).
\newblock Sex and species specific hearing mechanisms in mosquito flagellar
  ears.
\newblock \emph{Nature Communications} 9, paper number 3911.
\newblock \doi{10.1038/s41467-018-06388-7}
\bibAnnoteFile{modelseries}

\bibitem[{Tandina et~al.(2018)Tandina, Doumbo, Yaro, Traoré, Parola, and
  Robert}]{species}
Tandina, F., Doumbo, O., Yaro, A.~S., Traoré, S.~F., Parola, P., and Robert,
  V. (2018).
\newblock Mosquitoes (diptera: Culicidae) and mosquito-borne diseases in mali,
  west africa.
\newblock \emph{Parasites \& Vectors} 11, article number 467.
\newblock \doi{10.1186/s13071-018-3045-8}
\bibAnnoteFile{species}

\bibitem[{Tischner(1953)}]{tischner}
Tischner, H. (1953).
\newblock Über den {G}ehörsinn von {S}techmücken.
\newblock \emph{Acustica} 3, 335--343
\bibAnnoteFile{tischner}

\bibitem[{{van Noorden}(2014)}]{noorden}
{van Noorden}, L. (2014).
\newblock Auto-correlation and entrainment in the synchronous reproduction of
  musical pulse: Developments in childhood.
\newblock \emph{Procedia -- Social and Behavioral Sciences} 126, 117--118.
\newblock \doi{https://doi.org/10.1016/j.sbspro.2014.02.336}.
\newblock International Conference on Timing and Time Perception, 31 March –
  3 April 2014, Corfu, Greece
\bibAnnoteFile{noorden}

\bibitem[{Vasconcelos et~al.(2019)Vasconcelos, Nunes, Ribeiro, Prandi, and
  Rogers}]{dinarte}
Vasconcelos, D., Nunes, N., Ribeiro, M., Prandi, C., and Rogers, A. (2019).
\newblock Locomobis: a low-cost acoustic-based sensing system to monitor and
  classify mosquitoes.
\newblock In \emph{16th IEEE Annual Consumer Communications \& Networking
  Conference (CCNC)} (Las Vegas, NV, USA), 1--6.
\newblock \doi{10.1109/CCNC.2019.8651767}
\bibAnnoteFile{dinarte}

\bibitem[{Vasconcelos et~al.(2021)Vasconcelos, Yin, Wetjen, Herbst, Ziemer,
  F\"{o}rster et~al.}]{goodit}
Vasconcelos, D., Yin, M.~S., Wetjen, F., Herbst, A., Ziemer, T., F\"{o}rster,
  A., et~al. (2021).
\newblock Counting mosquitoes in the wild: An internet of things approach.
\newblock In \emph{Proceedings of the Conference on Information Technology for
  Social Good} (New York, NY, USA: Association for Computing Machinery), GoodIT
  '21, 43--48.
\newblock \doi{10.1145/3462203.3475914}
\bibAnnoteFile{goodit}

\bibitem[{Warren et~al.(2010)Warren, Lukashkin, and Russell}]{spontaneous}
Warren, B., Lukashkin, A.~N., and Russell, I.~J. (2010).
\newblock The dynein–tubulin motor powers active oscillations and
  amplification in the hearing organ of the mosquito.
\newblock \emph{Proc Biol Sci.} 277, 1761--1769.
\newblock \doi{10.1098/rspb.2009.2355}
\bibAnnoteFile{spontaneous}

\bibitem[{Warren and Russell(2011)}]{tunedis}
Warren, B. and Russell, I. (2011).
\newblock Mosquitoes on the wing “tune in” to acoustic distortion.
\newblock \emph{AIP Conference Proceedings} 1403, 479--480.
\newblock \doi{10.1063/1.3658134}
\bibAnnoteFile{tunedis}

\bibitem[{Windmill et~al.(2017)Windmill, Jackson, Pook, and
  Robert}]{invivowindmill}
Windmill, J. F.~C., Jackson, J.~C., Pook, V.~G., and Robert, D. (2017).
\newblock Frequency doubling by active in vivo motility of mechanosensory
  neurons in the mosquito ear.
\newblock \emph{Royal Society Open Science} 5, paper number 171082.
\newblock \doi{10.1098/rsos.171082}
\bibAnnoteFile{invivowindmill}

\bibitem[{Wishart et~al.(1962)Wishart, {van Sickle}, and Riordan}]{wishart}
Wishart, G., {van Sickle}, G.~R., and Riordan, D.~F. (1962).
\newblock Orientation of the males of aedes aegypti (l.) (diptera: Culicidae)
  to sound.
\newblock \emph{Can. Entom.} 94, 613--626.
\newblock \doi{10.4039/Ent94613-6}
\bibAnnoteFile{wishart}

\bibitem[{{World Health Organization}(2003)}]{surveillance}
{World Health Organization} (2003).
\newblock \emph{Guidelines for Dengue Surveillance and Mosquito Control}
  (Manila: WHO Library Cataloguing in Publication Data)
\bibAnnoteFile{surveillance}

\bibitem[{{World Health Organization}(2020)}]{who}
{World Health Organization} (2020).
\newblock Evaluation of genetically modified mosquitoes for the control of
  vector-borne diseases.
\newblock \emph{\url{https://www.who.int/publications/i/item/9789240013155},
  accessed: 2020-12-14}
\bibAnnoteFile{who}

\bibitem[{Yang et~al.(2012)Yang, Long, and Wang}]{2012-HarmonicSynchronization}
Yang, N., Long, Z., and Wang, F. (2012).
\newblock Harmonic synchronization model of the mating dengue vector
  mosquitoes.
\newblock \emph{Chinese Science Bulletin} 57.
\newblock \doi{10.1007/s11434-012-5445-z}
\bibAnnoteFile{2012-HarmonicSynchronization}

\bibitem[{Yin et~al.(2021)Yin, Haddawy, Nirandmongkol, Kongthaworn,
  Chaisumritchoke, Supratak et~al.}]{lyn}
Yin, M.~S., Haddawy, P., Nirandmongkol, B., Kongthaworn, T., Chaisumritchoke,
  C., Supratak, A., et~al. (2021).
\newblock A lightweight deep learning approach to mosquito classification from
  wingbeat sounds.
\newblock In \emph{Proceedings of the Conference on Information Technology for
  Social Good} (New York, NY, USA: Association for Computing Machinery), GoodIT
  '21, 37--42.
\newblock \doi{10.1145/3462203.3475908}
\bibAnnoteFile{lyn}

\bibitem[{Ziemer and Hinse(2017)}]{ice}
Ziemer, G. and Hinse, P. (2017).
\newblock Relation of maximum structural velocity and ice drift speed during
  frequency lock-in.
\newblock In \emph{24th International Conference on Port and Ocean Engineering
  under Arctic Conditions (POAC)} (Busan, Korea), paper number POAC17--071
\bibAnnoteFile{ice}

\bibitem[{Ziemer(2020)}]{book}
Ziemer, T. (2020).
\newblock \emph{Psychoacoustic Music Sound Field Synthesis}.
\newblock No.~7 in Current Research in Systematic Musicology (Cham: Springer).
\newblock \doi{10.1007/978-3-030-23033-3}
\bibAnnoteFile{book}

\bibitem[{Ziemer and Bader(2016)}]{complexp}
Ziemer, T. and Bader, R. (2016).
\newblock Complex point source model to calculate the sound field radiated from
  musical instruments.
\newblock \emph{Proceedings of Meetings on Acoustics} 25, 035001.
\newblock \doi{10.1121/2.0000122}
\bibAnnoteFile{complexp}

\bibitem[{Ziemer et~al.(2020)Ziemer, Koch, Sa-Ngamuang, Yin, Siai, Berkhausen
  et~al.}]{jasaabstract}
Ziemer, T., Koch, J., Sa-Ngamuang, C., Yin, M.~S., Siai, M., Berkhausen, B.,
  et~al. (2020).
\newblock A bio-inspired acoustic detector of mosquito sex and species.
\newblock \emph{The Journal of the Acoustical Society of America} 148,
  2480--2480.
\newblock \doi{10.1121/1.5146873}
\bibAnnoteFile{jasaabstract}

\end{thebibliography}





\end{document}